\documentclass[prd,preprint,tightenlines,floatfix,showpacs,preprintnumbers,nofootinbib,eqsecnum]{revtex4}

\usepackage[dvips,final]{graphicx}
\usepackage{amssymb}
\usepackage{amsmath}
\usepackage{amsfonts}
\usepackage{epsfig}
\usepackage{bm} 

\begin{document}
	
	
	\title{Unruh effect universality:  emergent conical geometry from density operator}
	\author{Georgy Y. Prokhorov$^{1,2}$}
	\email{prokhorov@theor.jinr.ru}
	
	\author{Oleg V. Teryaev$^{1,2,3}$}
	\email{teryaev@jinr.ru}
	
	\author{Valentin I. Zakharov$^{2,4,5}$}
	\email{vzakharov@itep.ru}
	\affiliation{
		{$^1$\sl
			Joint Institute for Nuclear Research, 141980 Dubna,  Russia \\
			$^2$ \sl
			Institute of Theoretical and Experimental Physics, NRC Kurchatov Institute,
			117218 Moscow, Russia
			\\
			$^3$ \sl
			National Research Nuclear University MEPhI (Moscow Engineering Physics Institute), Kashirskoe Shosse 31, 115409 Moscow, Russia
			\\
			$^4$ \sl
			School of Biomedicine, Far Eastern Federal University, 690950 Vladivostok, Russia\\
			$^5$ \sl
			Moscow Institute of Physics and Technology, 141700 Dolgoprudny, Russia
		}\vspace{1.5cm}
	}
	
	\begin{abstract}
		\vspace{0.5cm}
		The Unruh effect has been investigated from the point of view of the quantum statistical Zubarev density operator in space with the Minkowski metric. Quantum corrections of the fourth order in acceleration to the energy-momentum tensor of real and complex scalar fields, and Dirac field are calculated. Both massless and massive fields are considered. The method for regularization of discovered infrared divergences for scalar fields is proposed. The calculated corrections make it possible to substantiate the Unruh effect from the point of view of the statistical approach, and to explicitly show its universality for various quantum field theories of massless and massive fields. The obtained results exactly coincide with the ones obtained earlier by calculation of the vacuum average of energy-momentum tensor in a space with a conical singularity. Thus, the duality of two methods for describing an accelerated medium is substantiated. One may also speak about the emergence of geometry with conical singularity
		from thermodynamics. In particular, the polynomiality of the energy-momentum tensor and the absence of higher-order corrections in acceleration can be explicitly demonstrated.
	\end{abstract}
	
	\maketitle

\section{Introduction}
\label{sec:intro}

The properties of a medium with acceleration and vorticity are controlled by fundamental laws that arise at the quantum-field level \cite{Kharzeev:2012ph, Son:2009tf, Sadofyev:2010is, Stone:2018zel, Unruh:1976db, Vilenkin:1980zv}. Various quantum-field effects associated with acceleration and vorticity have been discovered: the chiral vortical effect (CVE)  \cite{Kharzeev:2012ph,Son:2009tf,Sadofyev:2010is,Stone:2018zel, Vilenkin:1980zv}, the Unruh effect \cite{Unruh:1976db}, phase transitions due to rotation \cite{Jiang:2016wvv} and acceleration \cite{Castorina:2012yg, Takeuchi:2015nga, Ohsaku:2004rv} of the medium. These effects are now also the subject of an experimental search in heavy ion collisions and quark-gluon plasma, in particular, vorticity, or more precisely the thermal vorticity tensor, can lead to polarization of hadrons \cite{Rogachevsky:2010ys, Florkowski:2018fap, Becattini:2019ntv, Baznat:2017jfj}, and acceleration is considered as a possible source of thermalization and hadronization \cite{Becattini:2008tx, Prokhorov:2019hif}.

A remarkable observation is that these effects allow to show the duality of various theoretical approaches. An example is the CVE, which can be obtained in the framework of different statistical approaches \cite{Buzzegoli:2017cqy, Prokhorov:2018bql,Prokhorov:2018qhq,Buzzegoli:2018wpy, Gao:2012ix}, using the hydrodynamic approach with the quantum axial anomaly \cite{Son:2009tf}, in the framework of an effective field theory \cite{Sadofyev:2010is}, on the basis of the gravitational anomaly \cite{Stone:2018zel}. Another example is the chiral magnetic effect \cite{Fukushima:2008xe, Sadofyev:2010is}.

In this paper, we focus on another well-known effect related to the motion of a medium such as the Unruh effect \cite{Unruh:1976db}.  According to this effect, the accelerated observer perceives the Minkowski vacuum as a medium with a temperature $ T_U $, proportional to the acceleration modulus  $ |a| $
\begin{eqnarray}
T_U=\frac{|a|}{2\pi}\,.
\label{TU}
\end{eqnarray}
This temperature is known as Unruh temperature.

In recent years, a fundamental theoretical approach has been developed that allows one to study the effects associated with acceleration. It is based on the Zubarev quantum statistical density operator \cite{Zubarev, Buzzegoli:2017cqy,Weert, Buzzegoli:2018wpy,Becattini:2015nva,Becattini:2019dxo}. Acceleration-related effects can be obtained by calculating quantum corrections to the mean values of the observed quantities in the inertial system described by the Minkowski metric. Thus, in particular, corrections up to the fourth order in acceleration to the energy-momentum tensor of Dirac field \cite{Prokhorov:2019cik, Prokhorov:2019hif}, as up to the second order for scalar field \cite{Buzzegoli:2017cqy, Becattini:2015nva, Buzzegoli:2018wpy} were calculated.

The mean values calculated in this way correspond to the normalization at which the contribution corresponding to the Minkowski vacuum is subtracted. Thus, it becomes possible to verify the Unruh effect from the point of view of quantum statistical mechanics. Indeed, this effect should lead to a situation where the mean values vanish at the proper temperature equal to the Unruh temperature: so the Minkowski vacuum corresponds to an accelerated medium with Unruh temperature \cite{Becattini:2017ljh, Dowker:1994fi,Frolov:1987dz, Florkowski:2018myy}.

Some results in this direction have already been obtained: in particular, the Unruh effect in this way was shown for the massless Dirac field \cite{Prokhorov:2019cik, Prokhorov:2019hif}, for which it was necessary to calculate fourth-order corrections in acceleration. In addition, a new effect was discovered related to the instability at Unruh temperature \cite{Prokhorov:2019hif, Becattini:2017ljh}.

In this paper, we test the statement about the Unruh effect in a statistical approach for a wider class of theories and show that it is general. We calculated the corrections to the energy-momentum tensor of the real and complex free massless scalar fields, as well as the quadratic mass corrections for scalar and Dirac fields. We show that in all investigated cases the calculated coefficients exactly satisfy the condition of vanishing at Unruh temperature. Thus, we conclude that the effect previously shown in \cite{Prokhorov:2019cik} is universal.

Despite the fact that the final conclusion about the Unruh effect is the same for fermions and scalar fields, the quantum corrections themselves have a different structure. If for fermions the fourth-order terms turned out to be finite, then for scalar fields these terms contain infrared divergences. We show that the appearance of infrared divergences in higher orders of perturbation theory follows from the general structure of the correlator, related to the coefficient of the $ n $-th order. We discuss the connection of infrared divergences with Matsubara zero modes and substantiate the procedure for their regularization.

An amazing observation that we make is the emergence of conical geometry in the Zubarev statistical approach.
This conclusion is based on the fact that the calculated quantum corrections exactly correspond to the results obtained in the framework of quantum field theory in a space with a conical singularity \cite{Dowker:1994fi, Dowker:1987pk, Iellici:1998ce, Bezerra:2006nu, Frolov:1987dz, Iellici:1997ud}. The emergence of conical geometry in the statistical approach was first noted by us and discussed in the case of massless fermions in \cite{Prokhorov:2019hif}. In this paper, we show that this is a general phenomenon, which is also true for massless scalar fields and for massive scalar and Dirac fields.

In particular, this duality of two approaches to describing the thermodynamics of an accelerated medium - statistical one, with the Zubarev operator, and geometrical approach with a conical singularity - allows one to obtain effects for an accelerated medium from the point of view of an inertial observer, using the available results for cosmic strings and vice versa \cite{Dowker:1994fi}.

The revealed duality also allows us to make a number of valuable statements regarding the statistical approach. First, in the conical singularity approach, the polynomiality of the energy-momentum tensor and the absence of higher-order corrections in acceleration can be explicitly demonstrated. Accordingly, we predict, on the basis of the duality, the absence of such corrections in the statistical approach. We have verified this statement directly in a particular case when calculating acceleration corrections for massive fields.

In addition, we get additional confirmation of the possibility of using the method of regularization of infrared divergences.

The paper has the following structure. In Section \ref{sec:intro2}, we discuss the possibility of describing an accelerated medium from the point of view of two different approaches, statistical and geometrical, based on field theory in conical space, and also discuss some aspects related to the discovered duality between them.
Section \ref{sec:real} provides calculation of quantum corrections of the fourth-order in acceleration to energy-momentum tensor of real scalar field in the massless case and in the quadratic order in mass. It is shown that when calculating acceleration corrections, infrared divergences arise and the procedure for their regularization is described. In the Section \ref{sec:complex} we give similar results for complex scalar field. Section \ref{sec:fermi} contains earlier results for massless fermions, as well as the calculation of quadratic mass corrections for massive Dirac field. Section \ref{sec:dual} discusses the emergence of conical geometry in the Zubarev statistical approach. In the Section \ref{sec:disc2} we establish the need for perturbative consideration of the effects of mass. Section \ref{sec:disc1} is devoted to various aspects of infrared divergences: their relationship with Matsubara zero modes, the need for summation in the massive case and the possibility of subtraction in the massless one. In the Section \ref{sec:disc} generalizing remarks about the observed universality of the Unruh effect in the statistical approach are made, and the consequences from the duality are discussed. In the Section \ref{sec:concl} conclusions are given.

\section{Overview of the duality between quantum-statistical and field-theoretical approaches}
\label{sec:intro2}

There exist two different approaches to evaluate various matrix elements,
in equilibrium
as functions of temperature $T$, chemical potential
$\mu$, vorticity $\omega_{\mu}$, acceleration $a_{\mu}$.
In the first approach the central role is played by Zubarev density operator $\hat{\rho}$ \cite{Zubarev, Buzzegoli:2017cqy,Weert}
\begin{equation}\label{two1}
\hat{\rho}~=~\frac{1}{Z}\exp\Big(-\beta_{\mu}\hat{P}^{\mu}+
\frac{1}{2}\varpi_{\mu\nu}\hat{J}^{\mu\nu}_x+\xi \hat{Q}\Big)\,,
\end{equation}
where $\beta_{\mu}=u_{\mu}/T$ and $u_{\mu}$ is the 4-velocity
of an element of the medium, $\hat{P}^{\mu}$ is the operator of the 4-momentum,
$\hat{J}^{\mu\nu}_x$
are the generators of the Lorentz transformations, shifted by vector $ x^{\mu} $,  $\hat{Q}$ is the charge
operator conjugated with the chemical potantial $\mu$.
Moreover
\begin{equation}\label{two2}
\varpi_{\mu\nu}~=~\frac{1}{2}\big(\partial_{\nu}\beta_{\mu}-
\partial_{\mu}\beta_{\nu}\big)\,,
\end{equation}
denotes the thermal vorticity.
In the hydrodynamic setup, the operator $\hat{J}^{\mu\nu}$ and tensor $ \varpi_{\mu\nu} $ in Eq. (\ref{two1})
can be decomposed as
\begin{eqnarray}
\hat{J}^{\mu\nu}&=&u^{\mu}\hat{K}^{\nu}-u^{\nu}\hat{K}^{\mu}-
\epsilon^{\mu\nu\rho\sigma}u_{\rho}\hat{J}_{\sigma}\,,
\nonumber \\
\varpi^{\mu\nu}&=&\alpha^{\mu}u^{\nu}-\alpha^{\nu}u^{\mu}-\epsilon^{\mu\nu\alpha\beta}u_{\alpha}w_{\beta}\,,
\end{eqnarray}
where $\hat{K}^{\mu}$ is the boost operator and $\hat{J}^{\mu}$ is the angular momentum, and vectors $ w_{\mu} $ and $ \alpha_{\mu} $ are related to the usual vorticity and acceleration by the relations $ w_{\mu}=\omega_{\mu}/T $ and  $ \alpha_{\mu}=a_{\mu}/T $ in the state of global equilibrium.

Note that in many cases, see, e.g., textbook \cite{landau} the term containing
the boost operator is not included into the density operator $\hat{\rho}$.
The main advantage and motivation to still use Eq. (\ref{two1}) is that it is explicitly
Lorentz-covariant, see, in particular \cite{Becattini:2017ljh}.
Note that inclusion of the boost operator makes explicit calculations
of the statistical averages
much more involved
since the boost operator does not commute with the Hamiltonian.

Statistically averaged matrix elements can be calculated perturbatively, by expanding the density operator
in $a_{\mu}, \omega_{\mu}$ and are expressed through the integrals over the momentum $d^3p$
with the weight function representing the Bose or Fermi distribution,
depending on the spin of the fields. In this way one evaluates one-loop effects
perturbatively within the statistical approach. It is worth emphasizing that
the expansion in acceleration of the equilibrium matrix elements is a novel
technique elaborated
only recently \cite{Becattini:2017ljh}. The most advanced explicit example of
such calculations at the moment is provided by the evaluation of terms of fourth order
in acceleration in expansion of the energy-momentum tensor \cite{Prokhorov:2019sss}.

The other approach can be called geometrical, or field theoretical.
It has been elaborated not in all the generality but only in some particular cases.
The most interesting case is $\mu, \omega =0, a\neq 0$.
It is well known that for an accelerated observer there exists a horizon.
In other words, the relevant space-time has a boundary. In the Euclidean version,
the temperature also acquires a geometrical meaning. Namely, the Euclidean time coordinate
is compact and the length of the corresponding circle is the inverse
temperature. Qualitatively, these simple  observations alone
allow for a geometrical treatment of the effects of acceleration,

Quantitatively, to accommodate temperature and acceleration as independent
parameters one considers
the Rindler space
with a conical singularity.  Introduce first
Rindler space with  Minkowskian signature
\begin{equation}\label{three}
ds^2~=~-r^2d\eta^2+dr^2+d{\bf x}_{\perp}^2\,,
\end{equation}
where $\eta=\gamma\lambda,\, x=r \cosh \eta,\,  t=r \sinh \eta,\,
d{\bf x}_{\perp}^2=dy^2+dz^2$ and $\gamma= \rm{const}$.
The proper acceleration $|a|=\sqrt{-a_{\mu}a^{\mu}}$ and the proper time of accelerated
observer  can be expressed in terms of the variables $\lambda$ and $r$
\begin{equation}\label{five}
|a|=r^{-1},~~~\tau=\gamma r\lambda\,.
\end{equation}
At finite temperature, the proper time $\tau$  is
to be made imaginary and periodic in the inverse proper temperature $T^{-1}$.
Thus, the metric (\ref{five}) takes the form
\begin{equation}\label{six}
ds^2=r^2d\eta^2+dr^2+d{\bf x}_{\perp}^2\,.
\end{equation}
The space (\ref{six})
describes a flat two-dimensional cone with angular deficit
$2\pi -|a|/T$ (plus a factorized transverse 2d Euclidean space).
Further details (and a picture) can be found in
\cite{Prokhorov:2019hif}.

Now, by expanding in modes one can evaluate energy density $\rho\equiv <T_0^0>$
associated with various quantum fields living on the space (\ref{six}) .
For massless particles of spin 1/2 the result is \cite{Dowker:1994fi}
\begin{equation}\label{seven}
\rho_{s=1/2}=\frac{7\nu^4}{960\pi^2r^4}+\frac{\nu^2}{96\pi^2r^4}-\frac{17}{960\pi^2r^4}\,,
\end{equation}
where $\nu~=~2\pi T r$. Since $ r=1/|a| $, Eq. (\ref{seven}) allows to unambiguously fix
the energy density, associated with quantum
massless field of spin 1/2 as function of  temperature and acceleration.

Thus, knowing the Eq. (\ref{seven}) one could have predicted in advance
the corresponding statistical one-loop average of the operator $\hat{T}_0^0$
evaluated along the lines
discussed above. There is a word of caution, however. There are two different
field theories behind derivation of Eq (\ref{seven}) and the statistical perturbation theory.
Any field-theoretic calculation might suffer uncertainties due to divergences.
Moreover, Eq. (\ref{seven}) is derived within so to say fundamental field theory
(on a nontrivial manifold) which is valid in ultraviolet. The statistical perturbation theory
deals with effective interaction introduced within thermodynamical approach rather
on a macroscopical scale, or in the infrared. Thus, the structure of divergences (if any)
could be very different. Moreover, there are specific features of the two theories
which might make the comparison of the results difficult. In particular, the first term
in the r.h.s. of Eq. (\ref{seven}) represents the pure thermal energy, and is fixed by the
Stephan-Boltzmann law. However, the last term in the r.h.s. of Eq. (\ref{seven})
represents the pure quantum Casimir effect due to existence of the boundary,
or horizon on the Rindler space. The statistical approach
weighs all the types of energy with the $1/T$ factor and, in this sense, deals with
various types of thermal contributions.

Within the statistical approach $\rho_{s=1/2}(a,T)$
has been evaluated perturbatively, see \cite{Prokhorov:2019hif}
and references therein, up to the fourth order in acceleration.
The result coincides with (\ref{seven}) exactly. Since the
statistical approach is formulated in Minkowski space and reproduces the Casimir
energy due to the Rindler horizon, one could even talk about the horizon
as an emergent notion.

In any case,  duality in description of $\rho_{s=1/2} (a,T)$ has been established without any
adjustment of parameters, or else.
Within the statistical approach the equilibrium energy is understood as energy
of real particles in the state with maximal entropy. While within the field-theoretic approach
the energy is a sum of the lowest energy levels of quantum oscillators, $(1/2)\omega_n$,
and is associated with virtual particles living on a non-trivial manifold.

Although the coincidence of the results is impressive by itself, there are a few subtle points
to be mentioned. First, turn to the structure of divergences.
The original expression for the Rindler vacuum energy (\ref{seven}) is ultraviolet divergent
and it is rendered finite  by subtracting the Minkowski expression, see \cite{Dowker:1994fi}.
Since the Minkowski vacuum is characterized by $T_U=|a|/2\pi$ one can say that
the vanishing of the r.h.s. of Eq (\ref{seven}) at the Unruh temperature
is imposed through a subtraction.
There is no need for such a subtraction in case of the statistical approach,
see \cite{Prokhorov:2019sss}.

It is worth also emphasizing that the Eq. (\ref{seven}) is valid only at $T \geq T_U$.
This is obvious in the geometric language, since the angular deficit cannot
be negative. On the other hand, if one concentrates on the statistical perturbation theory
in Minkowski space
there is no such a  limitation. Moreover, we are confronted with another question. Namely,
Eq. (\ref{seven}) makes it explicit that there are only three terms, of order $|a|^0, |a|^2, |a|^4$
in the expression for the Rindler vacuum energy. Explicit evaluation of such terms
within the statistical perturbation theory does
reproduce Eq. (\ref{seven}), for details see
\cite{Prokhorov:2019sss} and references therein. It remains unclear, however,
why higher terms
in the ratio $|a|^2/T^2$ are not in fact present.

Both problems have been resolved through working out a non-perturbative
expression for $\rho_{s=1/2}$ within the statistical approach. Namely,
the following integral
representation was proposed for $\rho_{s=1/2}$ in the massless case
\cite{Prokhorov:2019cik}
\begin{equation}\label{ten1}
\rho_{s=1/2}=~2\int\frac{d^3p}{(2\pi)^3}
\Big(\frac{|{\bf p}|+i|{a}|}{1+e^{\frac{|{\bf p}|}{T}+\frac{i|a|}{2T}}}+
\frac{|{\bf p}|-i|a|}{1+e^{\frac{|{\bf p}|}{T}-\frac{i|a|}{2T}}}
\Big) +4\int\frac{d^3 p}{(2\pi)^3}\frac{|{\bf p}|}{e^{\frac{2\pi |{\bf p}|}{|a|}}-1}\,.
\end{equation}
Eq. (\ref{ten1}) is motivated on theoretical grounds and goes through a number of
non-trivial checks, for details see \cite{Prokhorov:2019cik}
and references therein. In particular, expansion of
(\ref{ten1}) in the acceleration $|a|$ is exhausted by the first three terms and coincides
with (\ref{seven}), once the latter equation is rewritten in terms of $|a|$ and $T$.
Also, the representation (\ref{ten1}) reveals non-analyticity at the Unruh temperature,
$T_U=|a|/2\pi$ which cannot be detected perturbatively. Moreover, Eg (\ref{ten1})
suggests
an analytical continuation to $T< T_U$ and, in this respect, is more informative
than Eq. (\ref{seven}), for details see \cite{Prokhorov:2019hif}.

To summarize, the statistical perturbation theory and field theory on non-trivial manifolds
provide in case of massless fermions dual
descriptions which are rather complementary to each other
and allow  for novel applications.
In this paper  we explore the possibility of
extending these results to scalar fields and massive Dirac fields.

\section{Real scalar field}
\label{sec:real}

\subsection{Massless real scalar field}
\label{subsec:real1}

In the case of a medium with acceleration, the density operator (\ref{two1}) contains a term with acceleration and a boost operator
\begin{eqnarray}
\hat{\rho}=\frac{1}{Z}\exp\Big\{-\beta_{\mu}\hat{P}^{\mu}
-\alpha_{\mu} \hat{K}^{\mu}_x
\Big\} \,.
\label{operk}
\end{eqnarray}
Using  (\ref{operk}), corrections related to acceleration to the energy-momentum tensor can be calculated. From symmetry considerations it follows that in the fourth order of the perturbation theory the energy-momentum tensor has the form
\begin{eqnarray}
\langle \hat{T}^{\mu\nu}\rangle &=& (\rho_0 +A_1 T^2 |a|^2   +A_2 |a|^4 )u^{\mu}u^{\nu}-
(p_0+A_3 T^2 |a|^2  + A_4 |a|^4)\Delta^{\mu\nu} \nonumber \\
&&+(A_5 T^2+A_6 |a|^2)a^{\mu}a^{\nu}+\mathcal{O}(a^6) \qquad
\Delta^{\mu\nu}=g^{\mu\nu}-u^{\mu}u^{\nu}\,.
\label{EMTa4}
\end{eqnarray}

The purpose of this section is to calculate quantum corrections in acceleration to the mean value of energy-momentum tensor of free scalar field. The energy-momentum tensor of a free real scalar field has the well-known operator form
\begin{eqnarray}
\hat{T}^{\mu\nu} =\partial^{\mu}\hat{\varphi} \partial^{\nu}\hat{\varphi}-\frac{1}{2}g^{\mu\nu}(\partial_{\lambda}\hat{\varphi}\partial^{\lambda}\hat{\varphi}-m^2\hat{\varphi}^2)
\,.
\label{EMTrealFields}
\end{eqnarray}
In this subsection, we will consider the case of massless fields $m=0$.

The coefficients  $ \rho_0$ and $p_0 $ correspond to standard formulas for the energy density and pressure of gas of massless bosons (they can also be found directly using (\ref{operk}))
\begin{eqnarray}
\rho_0=\frac{\pi^2 T^4}{30}\,, \quad
p_0=\frac{\pi^2 T^4}{90}\,.
\label{zeroReal}
\end{eqnarray}
Coefficients $A_n$ are to be found on the basis of (\ref{operk}). In \cite{Buzzegoli:2017cqy, Becattini:2015nva}, the second-order coefficients in acceleration $A_1, A_3, A_5$ were calculated
\begin{eqnarray}
A_1^0=\frac{1}{12},\quad A_3^0=-\frac{1}{18},\quad
A_5^0=\frac{1}{12}\,,
\label{coef old real}
\end{eqnarray}
where the index 0 means that we are considering the case of $ m=0 $. However, in the second order, the condition necessary to justify the Unruh effect was not fulfilled. One can easily check using (\ref{zeroReal}) and (\ref{coef old real}), that $\langle \hat{T}^{\mu\nu}\rangle (T=T_U)\neq 0$. Therefore, higher order corrections must be calculated. We proceed to calculate the coefficients $ A_2, A_4, A_6 $.

The general algorithm of calculation of the hydrodynamic coefficients on the basis of (\ref{operk}) is described in \cite{Buzzegoli:2017cqy, Becattini:2015nva, Prokhorov:2018bql}, and in the case of fourth-order corrections, in \cite{Prokhorov:2019cik, Prokhorov:2019sss}. To find the coefficients $A_2, A_4, A_6$, it is necessary to expand Eq.~(\ref{operk}) in the series of the perturbation theory in the acceleration and calculate quantum correlators with boost operators and operator of the quantity under study. Rewriting the boost operator through the energy-momentum tensor, we will need to calculate five-point quantum correlators of the form
\begin{eqnarray}
&&C^{\alpha_1\alpha_2|\alpha_3\alpha_4
|\alpha_5\alpha_6|\alpha_7\alpha_8|\alpha_9\alpha_{10}|
ijkl} =\int_0^{|\beta|} d\tau_xd\tau_yd\tau_zd\tau_f d^3x d^3y d^3z d^3f \nonumber \\
&&\times x^{i} y^{j} z^{k} f^{l}\langle
T_{\tau} \hat{T}^{\alpha_1\alpha_2}(\tau_x,\bold{x})\hat{T}^{\alpha_3\alpha_4}(\tau_y,\bold{y})\hat{T}^{\alpha_5\alpha_6}(\tau_z,\bold{z})
\hat{T}^{\alpha_7\alpha_8}(\tau_f,\bold{f})\hat{T}^{\alpha_9\alpha_{10}}(0)
\rangle_{\beta(x),c}\,.
\label{C}
\end{eqnarray}
The index $\beta(x),c$ means that mean value is to be defined using the operator (\ref{operk}) with $\alpha^{\mu}=0$ and that only connected correlators are taken into account, $T_{\tau}$ means ordering by inverse temperature, and $|\beta|=\frac{1}{T}$. The coefficients in (\ref{EMTa4}) are expressed in terms of (\ref{C}) as follows
\begin{eqnarray} \nonumber
&&A_2 = \frac{1}{4!}C^{00|00|00|00|00|3333}\,,\quad A_4 = \frac{1}{4!}C^{00|00|00|00|33|2222}\,, \\
&&A_6 =-A_4+ \frac{1}{4!}C^{00|00|00|00|33|3333}
\,.
\label{AC}
\end{eqnarray}
We present the result of calculation for the coefficient $A_2$ in integral form
\begin{eqnarray}
A_2^0 &=& \int_0^{\infty} \frac{|{\bf p}|^3 d |{\bf p}|}{72\pi^2}\Big(n_B^{(4)}(|{\bf p}|)+|{\bf p}|\,n_B^{(5)}(|{\bf p}|)+\frac{7|{\bf p}|^2}{20} n_B^{(6)}(|{\bf p}|)+\frac{|{\bf p}|^3}{25} n_B^{(7)}(|{\bf p}|) \nonumber \\
&&+\frac{9|{\bf p}|^4}{5600} n_B^{(8)}(|{\bf p}|)\Big)
\,,
\label{A2Int}
\end{eqnarray}
where $n_B^{(k)}(p)=\frac{d^k}{d p^k}\frac{1}{e^{p/t}-1}$ is the derivative of the Bose-Einstein distribution.

An essential property of (\ref{A2Int}) is the appearance of infrared divergence in the limit $|{\bf p}|\to 0$. The appearance of this infrared divergence is a direct consequence of the Bose distribution pole for $p\to 0 $ in (\ref{A2Int}). It turns out that if we extract this divergence as a term of the Laurent series with negative power of the momentum, then the infinite integral (\ref{A2Int}) can be represented as the sum of the finite contribution and the divergent integral of the form $\int d|{\bf p}|/|{\bf p}|^2$.

A similar situation will be for the other coefficients in (\ref{EMTa4}). We write out the final expressions for the coefficients, representing them as the sum of the finite and diverging contributions
\begin{eqnarray}
A_2^0 &=& -\frac{11}{480\pi^2}+ \frac{4T}{15\pi^2}\int_0^{\infty}\frac{d|{\bf p}|}{|{\bf p}|^2}
\,, \nonumber \\
A_4^0 &=& \frac{19}{1440\pi^2}-\frac{6 T}{35\pi^2}\int_0^{\infty}\frac{d|{\bf p}|}{|{\bf p}|^2}\,, \nonumber \\
A_6^0 &=& -\frac{1}{48\pi^2}+\frac{26 T}{105\pi^2}\int_0^{\infty}\frac{d|{\bf p}|}{|{\bf p}|^2}\,.
\label{a2a4a6m0}
\end{eqnarray}

A standard technique in quantum theory of massless fields is to equate separately appearing divergent integrals of the form  $\int d|{\bf p}|/|{\bf p}|^2$ to zero, which is associated with the absence of a dimensional parameter in the integral \cite{Collins:1984xc,Pak:2010pt,Kleinert:2001ax}. This statement is known as Veltman's formula. Now we will follow this rule, while in the Section \ref{sec:disc1} we give a more detailed and rigorous justification.

Subtracting the divergences in (\ref{a2a4a6m0}) and  taking into account the coefficients (\ref{zeroReal}) and (\ref{coef old real}), we obtain the next expression for the energy-momentum tensor
\begin{eqnarray}
\langle \hat{T}^{\mu\nu}\rangle_{\mathrm{real}}^0 &=& \Big(\frac{\pi^2 T^4}{30} +\frac{T^2 |a|^2 }{12}   -\frac{11  |a|^4}{480\pi^2}\Big) u^{\mu}u^{\nu}-
\Big(\frac{\pi^2 T^4}{90}-\frac{T^2 |a|^2 }{18} \nonumber \\
&&+ \frac{19  |a|^4}{1440\pi^2}\Big) \Delta^{\mu\nu}+
 \Big(\frac{T^2}{12} -\frac{ |a|^2}{48 \pi^2}\Big) a^{\mu}a^{\nu}+\mathcal{O}(a^6)\,.
\label{tensorrealm0}
\end{eqnarray}
One can easily check that
$\langle \hat{T}^{\mu\nu}\rangle_{\mathrm{real}}^0$ vanishes at the Unruh temperature
\begin{eqnarray}
\langle \hat{T}^{\mu\nu}\rangle_{\mathrm{real}}^0  (T=T_U) = 0\,.
\label{UnruhReal}
\end{eqnarray}
The fulfilment of condition (\ref{UnruhReal}) is a direct indication of the Unruh effect: the Minkowski vacuum corresponds to the proper temperature equal to the Unruh temperature.

\subsection{Massive real scalar field}
\label{subsec:real2}

Mass corrections can be calculated using the same algorithm used in massless theory. Mass effects are contained in the energy-momentum tensor itself (\ref{EMTrealFields}), as well as in propagators \cite{Buzzegoli:2017cqy}. The terms of the order $m^2 T^2$ can be obtained from the standard formulas for $ \rho_0 $ and $ p_0 $, which describe the energy density and pressure of static gas of massive bosons. The corresponding standard formulas can also easily be calculated using the Zubarev density operator, and in this case it will be necessary to subtract the standard ultraviolet-diverging vacuum contribution. The finite parts will have the form
\begin{eqnarray}
\rho_0 &=& \frac{1}{2 \pi^2}\int_0^{\infty}d|{\bf p}|\,|{\bf p}|^2 E_p n_B(E_p)\,,\nonumber \\
p_0 &=& \frac{1}{6 \pi^2}\int_0^{\infty}\frac{d|{\bf p}|}{E_p}|{\bf p}|^4 n_B(E_p)
\,,
\label{coef zero m}
\end{eqnarray}
where $ E_p =\sqrt{m^2+|{\bf p}|^2} $. In the order $ m^2 $ we obtain, in particular, for $ \rho_0 $
\begin{eqnarray}
\rho_0^{m2} = \frac{\pi^2 T^4}{30}- \frac{m^2 T}{4 \pi^2}\int_0^{\infty}d|{\bf p}|\frac{1+ e^{|{\bf p}|/T}(|{\bf p}|/T-1)}{(e^{|{\bf p}|/T}-1)^2}\,.
\label{coef zero m 1}
\end{eqnarray}
The integral in (\ref{coef zero m 1}) converges and we obtain (similarly for $ p_0 $)
\begin{eqnarray}
\rho_0^{m2} &=& \frac{\pi^2 T^4}{30}- \frac{m^2 T^2}{24}\,, \nonumber \\ p_0^{m2} &=& \frac{\pi^2 T^4}{90}- \frac{m^2 T^2}{24}\,.
\label{coef zero m 2}
\end{eqnarray}

Corrections of the order $m^2 |a|^2$ can be obtained on the basis of the formulas \cite{Becattini:2015nva} or \cite{Buzzegoli:2017cqy}. In general case of $ m\neq 0 $ coefficients $ A_1, A_3, A_5 $ are
\begin{eqnarray}
A_1 &=& \frac{1}{48 \pi^2 T^2}\int_0^{\infty}d|{\bf p}|\,E_p(m^2+4 |{\bf p}|^2) n_B''(E_p)\,,\nonumber \\
A_3 &=& -\frac{1}{144 \pi^2 T^2}\int_0^{\infty}\frac{d|{\bf p}|}{E_p}|{\bf p}|^2(8|{\bf p}|^2+15 m^2 ) n_B''(E_p)
\,, \nonumber \\
A_5 &=& \frac{1}{24 \pi^2 T^2}\int_0^{\infty}\frac{d|{\bf p}|}{E_p}|{\bf p}|^2(2|{\bf p}|^2+3 m^2 ) n_B''(E_p)
\,.
\label{coef old 1}
\end{eqnarray}
We note, however, that we did not find these formulas for real scalar fields in the form (\ref{coef old 1}) in the literature. Exactly the same formulas, but with a factor 2, are given in \cite{Buzzegoli:2017cqy} for complex scalar fields. Despite the fact that it is obvious in advance that the result for a real scalar field will be two times smaller, we derived (\ref{coef old 1}) directly from the density operator with a scalar field, following the algorithm \cite{Buzzegoli:2017cqy}.

Corrections of the order $ m^2 $ can be obtained in the same way as above, by expanding the integrands in (\ref{coef old 1}) in a series in mass. In particular, for the coefficient $ A_1 $ we get
\begin{eqnarray} \nonumber
A_1^{m2} &=& \frac{1}{12}-\frac{m^2}{48 \pi ^2 T^5} \int_0^{\infty}d|{\bf p}|\, |{\bf p}| e^{|{\bf p}|/T} \Big(2 |{\bf p}| (4 e^{|{\bf p}|/T}+e^{2 |{\bf p}|/T}+1) \\
&&-3 T (e^{2 |{\bf p}|/T}-1)\Big)(e^{|{\bf p}|/T}-1)^{-4}\,.
\label{coef old 2}
\end{eqnarray}

As in the previous section, when calculating the terms $ |a|^4 $, the integral (\ref{coef old 2}) contains the infrared divergence of the form $ 1/|{\bf p}|^2 $. We isolate this divergence as a separate term. Then the coefficient $ A_1^{m2} $ will be presented as a combination of a finite term and a divergent integral. A similar situation is realized for the remaining coefficients in (\ref{coef old 1}). As a result, we obtain
\begin{eqnarray}
A_1^{m2} &=& \frac{1}{12}
+\frac{m^2}{96\pi^2 T^2}-\frac{m^2}{8 \pi ^2 T }\int_0^{\infty}\frac{d|{\bf p}|}{|{\bf p}|^2} \,,\nonumber \\
A_3^{m2} &=& -\frac{1}{18}+\frac{m^2}{96\pi^2 T^2}+\frac{m^2}{72 \pi ^2 T }\int_0^{\infty}\frac{d|{\bf p}|}{|{\bf p}|^2}
\,, \nonumber \\
A_5^{m2} &=& \frac{1}{12} -\frac{m^2}{12 \pi ^2 T} \int_0^{\infty}\frac{d|{\bf p}|}{|{\bf p}|^2}
\,.
\label{a1a3a5m2}
\end{eqnarray}

The terms of the order $ m^2 |a|^4/T^2 $ can be calculated on the basis of the formulas (\ref{AC}) and (\ref{C}), where it is necessary to keep the mass in the propagators and the energy-momentum tensor. Moreover, we find that these terms contain only infrared divergences, and the finite contribution in all coefficients is zero.

Accordingly, in the order $ m^2 $ we obtain (we keep the divergences that appeared earlier)
\begin{eqnarray}
A_2^{m2} &=& -\frac{11}{480\pi^2}+ \frac{4T}{15\pi^2}\int_0^{\infty}\frac{d|{\bf p}|}{|{\bf p}|^2}- \frac{2m^2T }{5\pi^2}\int_0^{\infty}\frac{d|{\bf p}|}{|{\bf p}|^4}
\,, \nonumber \\
A_4^{m2} &=& \frac{19}{1440\pi^2}-\frac{6 T}{35\pi^2}\int_0^{\infty}\frac{d|{\bf p}|}{|{\bf p}|^2}+ \frac{2m^2T }{35\pi^2}\int_0^{\infty}\frac{d|{\bf p}|}{|{\bf p}|^4}
\,, \nonumber \\
A_6^{m2} &=& -\frac{1}{48\pi^2}+\frac{26 T}{105\pi^2}\int_0^{\infty}\frac{d|{\bf p}|}{|{\bf p}|^2}- \frac{6m^2T }{35\pi^2}\int_0^{\infty}\frac{d|{\bf p}|}{|{\bf p}|^4}\,.
\label{a2a4a6m2}
\end{eqnarray}

Acting in the same way as in the case of infrared divergences in (\ref{a2a4a6m0}) and (\ref{a1a3a5m2}), we cancel the divergences $ \int d|{\bf p}|/|{\bf p}|^2 $ and $ \int d|{\bf p}|/|{\bf p}|^4 $ in (\ref{a2a4a6m2}). However, as will be shown in the Section \ref{sec:disc1}, the situation in the massive case becomes more non-trivial compared to the massless one.  In contrast to the massless case, it is now necessary to sum up the complete series of divergences $ \int d |{\bf p} | / | {\bf p} |^{2n} $ in all orders in mass. However, the result of such a summation does not contribute to the corrections of the order of $ m^0 $ and $ m^2 $, and therefore, effectively these divergences can be subtracted from the terms considered by us now. As a result, taking into account  (\ref{coef zero m 2}), (\ref{a1a3a5m2}) and (\ref{a2a4a6m2}) we obtain the following expression for the corrections of the order $m^2$ to the energy-momentum tensor
\begin{eqnarray}
\Delta\langle \hat{T}^{\mu\nu}\rangle_{\mathrm{real}}^{m2} &=& m^2\Big(-\frac{T^2}{24}+\frac{|a|^2}{96\pi^2} \Big) u^{\mu}u^{\nu} \nonumber \\
&& - m^2\Big(-\frac{T^2}{24}+\frac{|a|^2}{96\pi^2} \Big) \Delta^{\mu\nu}+\mathcal{O}(a^6)
\,.
\label{m2Scalar}
\end{eqnarray}
It follows from (\ref{m2Scalar}) that
\begin{eqnarray}
\Delta\langle \hat{T}^{\mu\nu}\rangle_{\mathrm{real}}^{m2} (T=T_U) = 0\,,
\label{tensorrealm2}
\end{eqnarray}
and thus, the condition necessary for the Unruh effect is also satisfied in the order $ m^2 $.

\section{Complex scalar field}
\label{sec:complex}

\subsection{Massless complex scalar field}
\label{subsec:complex1}

We will consider complex scalar fields with zero chemical potential $ \mu=0 $. The results obtained in this case are predictably the same as for the real scalar field, and differ only by a factor of 2, which is associated with a double number of degrees of freedom. However, at the technical level, the two cases are a bit different, which, in particular, is due to the fact that in the case of  complex scalar field there is an additional conjugate field. This simplifies the application of Wick theorem and from a technical point of view, calculations in the case of complex scalar fields turn out to be less complicated.

We start with the following standard expression for the energy-momentum tensor of complex scalar fields
\begin{eqnarray}
\hat{T}^{\mu\nu} =\partial^{\mu}\hat{\varphi}^{\dag} \partial^{\nu}\hat{\varphi}+\partial^{\nu}\hat{\varphi}^{\dag} \partial^{\mu}\hat{\varphi}-g^{\mu\nu}(\partial_{\lambda}\hat{\varphi}^{\dag}\partial^{\lambda}\hat{\varphi}-m^2\hat{\varphi}^{\dag}\hat{\varphi})
\,.
\label{EMTcomplex}
\end{eqnarray}
Calculating the corrections according to the algorithm described in the previous section, we find that all the coefficients for complex scalar fields are described exactly by the formulas (\ref{zeroReal}), (\ref{coef old real}) and (\ref{a2a4a6m0}), but with an additional coefficient of 2. And as a result, we get a 2 times larger value for the mean value of energy-momentum tensor
\begin{eqnarray}
\langle \hat{T}^{\mu\nu}\rangle_{\mathrm{complex}}^0=2 \langle \hat{T}^{\mu\nu}\rangle_{\mathrm{real}}^0
\,, \qquad \langle \hat{T}^{\mu\nu}\rangle_{\mathrm{complex}}^0(T=T_U)=0\,.
\label{tensorcomplexm0}
\end{eqnarray}
Thus, the Unruh effect is also observed statistically for massless complex scalar fields.

\subsection{Massive complex scalar field}
\label{subsec:complex2}

In the case of massive complex scalar fields, acceleration corrections can be calculated based on the formulas (\ref{C}) and (\ref{AC}) and the formulas from  \cite{Buzzegoli:2017cqy}, in exactly the same way as it was done in the previous section for real scalar fields. As a result, the mass corrections are described by the same formulas as for the real scalar fields (\ref{coef zero m 2}), (\ref{a1a3a5m2}) and (\ref{a2a4a6m2}), but with an additional coefficient of 2. The final result for the corrections of the order $ m^2 $ to the mean value of the energy-momentum tensor is the doubled real scalar one
\begin{eqnarray}
\Delta\langle \hat{T}^{\mu\nu}\rangle_{\mathrm{complex}}^{m2}=2 \Delta\langle \hat{T}^{\mu\nu}\rangle_{\mathrm{real}}^{m2}
\,, \qquad \Delta\langle \hat{T}^{\mu\nu}\rangle_{\mathrm{complex}}^{m2}(T=T_U)=0\,.
\label{tensorcomplexm2}
\end{eqnarray}
Thus, massive complex scalar field also satisfies the condition associated with the Unruh effect.

\section{Dirac field}
\label{sec:fermi}

\subsection{Massless Dirac field}
\label{subsec:fermi1}

In this section, for completeness, we present the results for massless fermions obtained in \cite{Prokhorov:2019cik}.
The calculation of the coefficients (\ref{EMTa4}) for fermions can be done on the basis of the formulas (\ref{C}) and (\ref{AC}), where it is necessary to use the energy-momentum tensor of fermions. As a result, we obtain the following coefficients
\begin{eqnarray}
&&\rho_0 = \frac{7 \pi^2 T^4}{60}\,, \quad
p_0 = \frac{7\pi^2 T^4}{180}\,, \quad
A_1^0 = \frac{1}{24}\,, \quad
A_2^0 = -\frac{17}{960 \pi^2}\,, \nonumber \\
&& A_3^0 = \frac{1}{72}\,, \quad
A_4^0 = -\frac{17}{2880 \pi^2}\,, \quad
A_5^0 = 0\,, \quad
A_6^0 = 0\,,
\label{coeffermim0}
\end{eqnarray}
where the second-order coefficients $ A_1, A_3, A_5 $ were calculated in \cite{Buzzegoli:2017cqy}. The energy-momentum tensor takes the form
\begin{eqnarray}
\langle \hat{T}^{\mu\nu}\rangle_{\mathrm{fermi}}^0 &=& \Big(\frac{7 \pi ^2 T^4}{60}+\frac{T^2 |a|^2}{24} -\frac{17 |a|^4}{960\pi^2} \Big) u^{\mu}u^{\nu}\nonumber \\
&&-
\Big(\frac{7 \pi ^2 T^4}{180}+\frac{T^2 |a|^2}{72} -\frac{17|a|^4}{2880\pi^2} \Big) \Delta^{\mu\nu}+\mathcal{O}(a^6)\,.
\label{tensorfermim0}
\end{eqnarray}
Unlike scalar fields (\ref{a2a4a6m0}), the energy-momentum tensor of fermions is free from infrared divergences. This situation corresponds to the fact that in the Fermi-Dirac distribution, in contrast to the Bose distribution, there is no singularity at zero energy.

It is easy to see that the Unruh effect is statistically fulfilled
\begin{eqnarray}
\langle \hat{T}^{\mu\nu}\rangle_{\mathrm{fermi}}^0(T=T_U)=0\,.
\label{UnruhFermi}
\end{eqnarray}

\subsection{Massive Dirac field}
\label{subsec:fermi2}

Coefficients of the order $ m^2 $ can be calculated in the same way as previously. As a result, we obtain
\begin{eqnarray}
&&\rho_0^{m2} = \frac{7 \pi^2 T^4}{60}-\frac{m^2 T^2}{12}\,, \quad
p_0^{m2} = \frac{7\pi^2 T^4}{180}-\frac{m^2 T^2}{12}\,, \quad
A_1^{m2} = \frac{1}{24}+\frac{m^2 }{48 \pi^2 T^2}\,, \nonumber \\
&&A_2^{m2} = -\frac{17}{960 \pi^2}\,, \quad A_3^{m2} = \frac{1}{72}+\frac{m^2 }{48 \pi^2 T^2}\,, \quad
A_4^{m2} = -\frac{17}{2880 \pi^2}\,, \nonumber \\
&& A_5^{m2} = 0\,, \quad
A_6^{m2} = 0\,,
\label{coeffermim2}
\end{eqnarray}
where the coefficients $ \rho_0^{m2}, p_0^{m2} $ are calculated by expanding the standard formulas for the energy and pressure of the static gas of massive fermions. The coefficients $ A_1^{m2}, A_3^{m2}, A_5^{m2} $ are calculated by expanding in mass the formulas from \cite{Buzzegoli:2017cqy}, while the coefficients $ A_2^{m2}, A_4^{m2}, A_6^{m2} $ can be obtained based on the formulas (\ref{C}) and (\ref{AC}), by holding the mass in the propagators and the energy-momentum tensor (complete formulas are not given because of their too large size). Taking into account (\ref{coeffermim2}), we obtain the next expression for the mean value of the energy-momentum tensor of Dirac field in the order $ m^2 $
\begin{eqnarray}
\langle \hat{T}^{\mu\nu}\rangle_{\mathrm{fermi}}^{m2} &=& \langle \hat{T}^{\mu\nu}\rangle_{\mathrm{fermi}}^0 + m^2\Big(-\frac{T^2}{12}+\frac{|a|^2}{48\pi^2} \Big) u^{\mu}u^{\nu}-
m^2\Big(-\frac{T^2}{12}+\frac{|a|^2}{48\pi^2}\Big)\Delta^{\mu\nu}\nonumber \\
&& +\mathcal{O}(a^6)
\,,
\label{tensorfermim2}
\end{eqnarray}
where $\langle \hat{T}^{\mu\nu}\rangle_{\mathrm{fermi}}^0$ corresponds to massless case (\ref{tensorfermim0}). It is easy to see that the energy-momentum tensor (\ref{tensorfermim2}) satisfies the condition resulting from the Unruh effect
\begin{eqnarray}
\langle \hat{T}^{\mu\nu}\rangle_{\mathrm{fermi}}^{m2} (T=T_U) = 0\,.
\label{UnruhFermim2}
\end{eqnarray}

According to (\ref{coeffermim2}), when calculating corrections with acceleration in the hydrodynamic coefficients of massive fermions, there are no divergences and all terms are finite. The same situation was in the case of massless fermions. It is also important to note that the terms of higher order $ m^2 |a|^4/T^2 $, as can be seen from the formulas for $ A_2^{m2}, A_4^{m2}, A_6^{m2} $, are equal to zero.

\section{The emergence of conical geometry in the Zubarev statistical approach}
\label{sec:dual}

It turns out that all the formulas (\ref{tensorrealm0}), (\ref{tensorrealm2}), (\ref{tensorcomplexm0}), (\ref{tensorcomplexm2}), (\ref{tensorfermim0}), (\ref{tensorfermim2}), calculated in the framework of the perturbation theory following from the Zubarev density operator (\ref{operk}) in an ordinary flat Minkowski space, can be obtained in another approach based on the consideration of a space with a conical singularity  \cite{Dowker:1994fi, Dowker:1987pk, Iellici:1998ce, Bezerra:2006nu, Frolov:1987dz, Iellici:1997ud}. This allows us to talk about the emergent conical geometry in the statistical approach of Zubarev or about the duality of the two approaches. The first indications of such duality were noted by us in \cite{Prokhorov:2019hif}, and now we will show it at a more general level.

Now, to consider the medium with acceleration, we move into space-time with an event horizon described by the Rindler metric (\ref{three}), which at finite temperature  takes the form (\ref{six}).
As noticed in the Section \ref{sec:intro2}, the space, described by (\ref{six}) contains a flat two-dimensional
cone with an angular deficit $2\pi- |a|/T $. One of the important properties of the
space (\ref{six}) is the presence of a conical singularity at $ r=0 $. We note in this case that despite the fact that the acceleration depends on the coordinate as $ | a | = 1/r $, the angular deficit is constant, since $ T \sim 1/r $, and $ | a | / T = \rm{ const} $, which is a consequence of global thermodynamic equilibrium \cite{Buzzegoli:2017cqy}.

In the papers \cite{Dowker:1994fi, Dowker:1987pk, Iellici:1998ce, Bezerra:2006nu, Frolov:1987dz, Iellici:1997ud} a quantum field theory in space-time of a cosmic string was considered. This space-time is equivalent to the Euclidean Rindler space-time (\ref{six}) up to change of the numbering of coordinates \cite{Dowker:1994fi}. In particular, the energy density in space-time (\ref{six}) turns out to be equal to the vacuum average of the component $ T_2^2 $ of the energy-momentum tensor  in space-time of the cosmic string: $ \rho_{Rindler}= \langle T_2^2 \rangle_{string}  $. In \cite{Dowker:1994fi, Dowker:1987pk, Iellici:1998ce, Bezerra:2006nu, Frolov:1987dz, Iellici:1997ud} the expressions for the $ \langle T_2^2 \rangle $ of massless real scalar field ($ s=0 $) and fermion field ($ s=1/2 $) are given
\begin{eqnarray}
\langle T_2^2 \rangle_{s=0} &=& \frac{\nu^4}{480 \pi^2 r^4} +\frac{\nu^2 }{48 \pi^2 r^4}   -\frac{11}{480\pi^2 r^4}\,, \nonumber \\
\langle T_2^2 \rangle_{s=1/2} &=& \frac{7\nu^4}{960 \pi^2 r^4} +\frac{\nu^2 }{96 \pi^2 r^4}   -\frac{17}{960\pi^2 r^4}
\,,
\label{t22conm0}
\end{eqnarray}
where  $ \nu $ defines the period $ 2\pi/\nu $ of the angular coordinate for the space-time of the cosmic string. For $ \nu=1 $, the conical singularity disappears. In order to move to space (\ref{six}), it is necessary to replace $ \nu = 2\pi T r $ and take into account the relation $ \rho_{Rindler}= \langle T_2^2 \rangle_{string}  $. As a result, we get
\begin{eqnarray}
\rho_{s=0} &=& \frac{\pi^2 T^4}{30} +\frac{T^2 }{12 r^2}   -\frac{11}{480\pi^2 r^4}\,, \nonumber \\
\rho_{s=1/2} &=& \frac{7\pi^2 T^4}{60} +\frac{T^2 }{24 r^2}   -\frac{17}{960\pi^2 r^4}
\,.
\label{enconm0r}
\end{eqnarray}
Taking into account (\ref{five}) we get exactly the energy density from (\ref{tensorrealm0}) and (\ref{tensorfermim0})
\begin{eqnarray}
\rho_{s=0} &=& \frac{\pi^2 T^4}{30} +\frac{T^2 |a|^2 }{12}   -\frac{11 |a|^4}{480\pi^2}\,, \nonumber \\
\rho_{s=1/2} &=& \frac{7\pi^2 T^4}{60} +\frac{T^2 |a|^2}{24}   -\frac{17 |a|^4}{960\pi^2}
\,.
\label{enconm0a}
\end{eqnarray}

In \cite{Iellici:1998ce, Iellici:1997ud} the results of calculation of quantum corrections with quadratic mass corrections for real scalar fields are given
\begin{eqnarray}
\langle T_2^2 \rangle^{m2}_{s=0} &=& \frac{\nu^4}{480 \pi^2 r^4} +\frac{\nu^2 }{48 \pi^2 r^4}   -\frac{11}{480\pi^2 r^4}+\frac{m^2}{96 \pi^2 r^2}(1-\nu^2)\,,
\label{t22conm2scal}
\end{eqnarray}
which, taking into account the comments made earlier, leads to energy density in space (\ref{six})
\begin{eqnarray}
\rho^{m2}_{s=0} = \frac{\pi^2 T^4}{30} +\frac{T^2 |a|^2 }{12}   -\frac{11 |a|^4}{480\pi^2}+m^2\Big(-\frac{T^2}{24}+\frac{|a|^2}{96\pi^2}\Big)\,.
\label{enconm2scal}
\end{eqnarray}
This formula exactly coincides with the expression for energy density, following from (\ref{tensorfermim2}).

In \cite{Bezerra:2006nu} the case of massive fermions in a space of cosmic string is considered. However, we did not find a derivation of corrections of the order $ m^2 $ from the general formula given in \cite{Bezerra:2006nu}. Further we present such a calculation. According to \cite{Bezerra:2006nu} and taking into account the correspondence between the Rindler metric and the metric with a cosmic string, the energy density of fermions with mass $ m $ is described by the formula
\begin{eqnarray}
\rho_{s=1/2} = -3 T_0+T_1\,,
\label{bez1}
\end{eqnarray}
where $ T_0 $ and $ T_1 $ are of the form
\begin{eqnarray}
T_0 &=& \frac{|a| m^2 T \cos \frac{\pi ^2
   T}{|a|} }{\pi ^2 }\int_{0}^{\infty} \frac{ dy\, K_2\left(\frac{2 m \cosh
  \frac{y}{2}}{|a|}\right) \sinh \frac{y}{2} \sinh \frac{\pi  T y}{|a|} }{\cosh^2 \frac{y}{2}\left(\cosh \frac{2 \pi  T y}{|a|} - \cos \frac{2
   \pi ^2 T}{|a|}\right)}\,, \nonumber \\
T_1 &=& -\frac{2 m^3 T \cos \frac{\pi ^2
   T}{|a|} }{\pi^2 }\int_{0}^{\infty} \frac{ dy\, K_1\left(\frac{2 m \cosh
   \frac{y}{2}}{|a|}\right) \sinh \frac{y}{2} \sinh \frac{\pi  T y}{|a|} }{\cosh\frac{y}{2}\left(\cosh \frac{2 \pi  T y}{|a|} - \cos \frac{2
   \pi ^2 T}{|a|}\right)}\,,
\label{bez2}
\end{eqnarray}
where $ K_1 $ and $ K_2 $ are the modified Bessel functions of the second kind. We expand the functions in (\ref{bez2}) in a series in mass up to the order of $ m^2 $. Then the energy density will take the form
\begin{eqnarray}
\rho^{m2}_{s=1/2} =I_0+ m^2 I_1\,,
\label{bez3}
\end{eqnarray}
where the integrals $ I_0 $ and $ I_1 $ are of the form
\begin{eqnarray}
I_0 &=& -\frac{3 |a|^3 T \cos \frac{\pi ^2
   T}{|a|} }{2\pi ^2 }\int_{0}^{\infty} \frac{ dy\, \sinh \frac{y}{2} \sinh \frac{\pi  T y}{|a|} }{\cosh^4 \frac{y}{2}\left(\cosh \frac{2 \pi  T y}{|a|} - \cos \frac{2
   \pi ^2 T}{|a|}\right)}\,, \nonumber \\
I_1 &=& \frac{|a| T  \cos \frac{\pi ^2
   T}{|a|} }{2\pi ^2 }\int_{0}^{\infty} \frac{ dy\, \sinh \frac{y}{2} \sinh \frac{\pi  T y}{|a|} }{\cosh^2 \frac{y}{2}\left(\cosh \frac{2 \pi  T y}{|a|} - \cos \frac{2
   \pi ^2 T}{|a|}\right)}\,.
\label{bez4}
\end{eqnarray}
The integrals (\ref{bez4}), can be found
\begin{eqnarray}
I_0 &=& \frac{7 \pi ^2 T^4}{60}+\frac{T^2 |a|^2}{24} -\frac{17 |a|^4}{960\pi^2}\,, \nonumber \\
I_1 &=& -\frac{T^2}{12}+\frac{|a|^2}{48\pi^2} \,,
\label{bez5}
\end{eqnarray}
and, thus, the energy density for fermions in the order $ m^2 $  takes the following form
\begin{eqnarray}
\rho^{m2}_{s=1/2} = \frac{7 \pi ^2 T^4}{60}+\frac{T^2 |a|^2}{24} -\frac{17 |a|^4}{960\pi^2} + m^2\Big( -\frac{T^2}{12}+\frac{|a|^2}{48\pi^2}\Big)\,.
\label{bez6}
\end{eqnarray}
The formula (\ref{bez6}) exactly matches the result from the Zubarev operator  (\ref{tensorfermim2}).

Summarizing, despite the difference between the two methods, statistical (with Zubarev operator) and geometrical (either in Euclidean Rindler space-time or space-time of cosmic string), both at the ideological and technical levels, it is striking that both methods lead to exactly the same results: formulas (\ref{tensorrealm0}), (\ref{tensorrealm2}), (\ref{tensorfermim0}), (\ref{tensorfermim2}) exactly coincide to  (\ref{enconm0a}), (\ref{enconm2scal}) and (\ref{bez6}).

Thus, we should talk about the emergent conical geometry in the statistical approach of Zubarev or about the duality of the two methods. The first method is statistical; all calculations are carried out in a flat space described by the Minkowski metric. This method ``knows nothing'' about curvilinear coordinates, but nevertheless leads to relations directly following from the Unruh effect: the coefficients turn out to be precisely such that the Minkowski vacuum corresponds to a proper temperature equal to Unruh temperature. The second method considers space with a boundary, an event horizon, which then transforms into space with a conical singularity. In this method, it is obvious in advance that all observables vanish at the Unruh temperature, since even at the level of the Green function, subtraction was performed at the Unruh temperature. In this sense, the first approach is an independent verification of the Unruh effect, while the second considers it as the initial premise.

\section{Beyond expansion in mass: the need for perturbative consideration}
\label{sec:disc2}

After reading Sections \ref{subsec:real2}, \ref{subsec:complex2} and \ref{subsec:fermi2}, the question may arise - is it possible to evaluate the Unruh effect without using mass expansion? Indeed, the coefficients of $ |a|^2 $ and $ |a|^4 $ terms can be found outside the expansion in mass, by accounting mass in propagators. We confine ourselves to considering the energy density of massive fermions \footnote{We are grateful to F. Becattini and M. Buzzegoli, with whom we discussed the issues investigated in the next two sections.}.

In the zero order in acceleration, the energy density of Dirac fields is well known
\begin{eqnarray}
\rho_0 (m,T) &=& \frac{2}{ \pi^2}\int_0^{\infty}d|{\bf p}|\,|{\bf p}|^2 E_p n_F(E_p)
\,.
\label{rho0Dirac}
\end{eqnarray}
Coefficient $ A_1 (m,T) $ of $ |a|^2 $ term is described by the formula (5.30) in \cite{Buzzegoli:2017cqy}. Coefficient $ A_2 (m,T) $ was found in \cite{Prokhorov:2019sss} and has the form
\begin{eqnarray}
A_2 (m,T) &=& \int_0^{\infty} \frac{|{\bf p}|^2 d |{\bf p}|}{2880\pi^2 E_p}\Big(5(15 m^2+19|{\bf p}|^2)n_F^{(4)}(E_p)+8 E_p(9 m^2+20|{\bf p}|^2)n_F^{(5)}(E_p)\nonumber \\
&&+2(4 m^4+25 m^2  |{\bf p}|^2+27|{\bf p}|^4) n_F^{(6)}(E_p)+
\frac{16}{35}E_p|{\bf p}|^2(5 m^2 +14 |{\bf p}|^2) n_F^{(7)}(E_p) \nonumber \\
&&+\frac{9}{35}E_p^2 |{\bf p}|^4 n_F^{(8)}(E_p)\Big)
\,,
\label{A2IntD}
\end{eqnarray}
(formula (\ref{A2IntD}) is equivalent to that given in \cite{Prokhorov:2019sss}). As a result, we can check what the energy density is at Unruh temperature for an arbitrary mass, without using mass expansion
\begin{eqnarray}
\rho(m,T_U)=\rho_0(m, T_U)+A_1(m, T_U)T_U^2 |a|^2+A_2(m, T_U) |a|^4\,.
\label{rhomdir}
\end{eqnarray}

\begin{figure}[!h]
	\centering 
	\centerline{\includegraphics[width=0.7\textwidth]{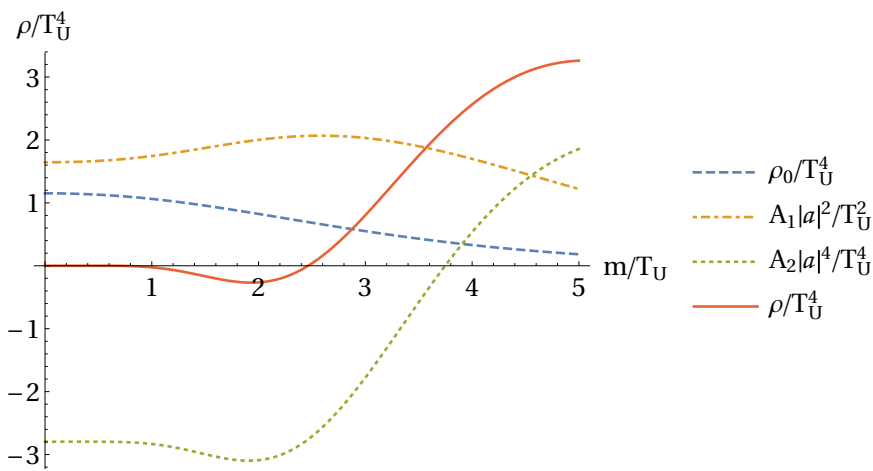}}
	\caption{\label{fig:1} Various contributions to the energy density in (\ref{rhomdir})  at Unruh temperature and their sum as functions of mass.}
\end{figure}

The graph of (\ref{rhomdir}) is shown in Fig.\ref{fig:1}. As might be expected, for small masses $ \rho(m,T_U) \approx 0$, however, for larger masses this condition is violated. This fact should not be surprising - in fact, in formula (\ref{rhomdir}) we have taken into account only terms up to the order $ |a|^4 $. Section \ref{sec:disc} gives arguments in favor of the absence of corrections to (\ref{rhomdir}), but only in the case of a mass equal to zero or in the corrections $ m^2 $. In general case, such a correction arises and should be taken into account when considering the Unruh effect. There is no doubt that in the general case the energy density should turn to zero at the Unruh temperature, since the Unruh effect is a universal phenomenon (look, in particular, (\ref{bez4})).

Given the comments made, the mass expansion used in the previous sections is seen as a suitable way of analyzing the Unruh effect in massive theories.

\section{Infrared divergences: summation, zero modes, nonperturbative terms}
\label{sec:disc1}

In this section, we discuss in more detail the infrared divergences that appeared in the Sections \ref{sec:real}, \ref{sec:complex}. The infrared divergences arose for both massless (\ref{a2a4a6m0}) and massive scalar fields (\ref{a1a3a5m2}), (\ref{a2a4a6m2}). We first consider the case with massless fields and restrict ourselves to the consideration of divergences in the energy density (the conclusions made below are to be valid for other divergences in the energy-momentum tensor).

First, it can be shown from the general structure of the perturbation theory that in $ 2n $-th order in acceleration, a leading infrared divergence of the form $ \int d|{\bf p}|/|{\bf p}|^{2n-2} $ arises, unless the coefficient at it turns out to be zero. Indeed, the contribution to the energy density of a massless real scalar fields in the $ 2n $-th order in acceleration can be schematically represented in the form  \cite{Prokhorov:2019sss, Buzzegoli:2017cqy}
\begin{eqnarray}
\rho\sim A_n |a|^{2n}&\sim& |a|^{2n}
\sum_{\omega_{1}...\omega_{2n+1}}\int d\tau_1...d\tau_{2n} d^3p_1...d^3p_{2n+1}d^3x_1...d^3x_{2n}x^{(3)}_1...x^{(3)}_{2n} \nonumber\\
&&\times \Delta(P_1)...\Delta(P_{2n+1}) e^{{\bf p x}}e^{\tau \omega}\prod_{i=1}^{2n+1}D^{00}_i\,,
\label{infr1}
\end{eqnarray}
where $ \Delta(P)=1/(P^2+m^2) $ is the scalar thermal Green function, $ {\bf p x} $ denotes a linear combination of scalar products of the form $ {\bf p}_i {\bf x}_j $, $ \tau \omega $ denotes a linear combination of products of the form $ \tau_i \omega_j $, and $ D^{00}_i $ denotes $ D^{00}(\pm i P_{i^{\prime}},\pm i P_{i^{\prime\prime}}) $ - the operator that occurs when the energy-momentum tensor is presented in a split form \cite{Prokhorov:2019sss, Buzzegoli:2017cqy}. Summing over the Matsubara frequencies and leaving in the expression (\ref{infr1}) only the terms contributing to the leading divergence, we obtain
\begin{eqnarray}
A_n |a|^{2n}&\sim& |a|^{2n}
\int d^3p \frac{\partial^{2n}}{(\partial p_i)^{2n}} \Big( \frac{1}{|{\bf p}_1|...|{\bf p}_{2n+1}|} \prod_{i=1}^{2n+1}D^{00}_i n_B(|{\bf p}_i|) \Big)\Big|_{\bf p_i=\pm \bf p}\,,
\label{infr2}
\end{eqnarray}
where $ \frac{\partial^{2n}}{(\partial p_i)^{2n}} $ denotes a linear combination of derivatives  of the $ 2n $-th order  of the form $ \frac{\partial^{2n}}{\partial p^{(3)}_{i_1}\partial p^{(3)}_{i_2}...\partial p^{(3)}_{i_{2n}}} $ (without proof, we assume that the explicit dependence on the coordinates in any order can be rewritten through the derivatives and the integration of the exponential factor $ e^{{\bf p x}} $ leads to the replacement of all the momenta of the form $ \bf p_i\to \pm \bf p $ as it was in all orders of perturbation theory considered so far).
Given that $ D^{00}_i $ is quadratic in momenta, the first term of the Laurent series, which gives the leading infrared divergence, in (\ref{infr2}) has the form
\begin{eqnarray}
A_n |a|^{2n}&\sim& |a|^{2n} T
\int d|{\bf p}|/|{\bf p}|^{2n-2}\,,
\label{infr3}
\end{eqnarray}
which was to be proved.

On the other hand, it can be shown that the diverging contribution in the massless case (\ref{a2a4a6m0}) exactly corresponds to the Matsubara zero modes. Indeed, expression (\ref{A2Int}) is formed from a set of correlators of the same type, the number of which is determined by Wick theorem (24 for Dirac fields and complex scalar fields and $24 \times 16$ for real scalar fields). In particular, for real scalar fields, the coefficient $ A_2 $ equals \cite{Prokhorov:2019sss}
\begin{eqnarray}\label{A2corrform}
&&A_2=\int \frac{d\tau_x d\tau_y d\tau_z d\tau_f d^3 x d^3 y d^3 z d^3 fd^3pd^3qd^3kd^3rd^3l}{4!(2\pi)^{15}}x^{3}y^{3}z^{3}f^{3} \\
&&\times e^{-i{\bf p}({\bf x}-{\bf y})-i{\bf q}{\bf x}-i{\bf k}({\bf y}-{\bf z})-i{\bf r}({\bf z}-{\bf f})-i{\bf l}{\bf f}}\nonumber  \\
&&
\times \sum_{p_n, q_n, k_n, r_n,l_n}\frac{1}{|\beta|^5} e^{ip_n(\tau_x -\tau_y)+iq_n\tau_x+ik_n (\tau_y-\tau_z) +ir_n(\tau_z-\tau_f)+il_n\tau_f}\nonumber \\
&&\times \Delta(P)\Delta(Q)\Delta(K) \Delta(R)\Delta(L)\nonumber \\
&&\times\mathcal{D}^{00}(iL ,-iR )
\mathcal{D}^{00}(iR ,-iK )
\mathcal{D}^{00}(iK ,-iP )
\mathcal{D}^{00}(iP ,iQ )\mathcal{D}^{00}(-iQ ,-iL )+....\,, \nonumber
\end{eqnarray}
where ellipsis includes similar $24 \times 16 - 1$ terms from other correlators in Wick theorem. Keeping in (\ref{A2corrform}) the contribution of only $ p_n=q_n=k_n=r_n=l_n=0  $ and making the necessary integrations, we find that this contribution entirely enters the infrared divergence
\begin{eqnarray}
A^{\omega=0}_2=\frac{4T}{384\times15\pi^2}\int_0^{\infty}\frac{d|{\bf p}|}{|{\bf p}|^2}+....\,.
\label{A2corrform1}
\end{eqnarray}
A direct calculation of the remaining correlators shows that their contributions are equal and, as a result, we obtain
\begin{eqnarray}
A^{\omega=0}_2=\frac{4T}{15\pi^2}\int_0^{\infty}\frac{d|{\bf p}|}{|{\bf p}|^2}\,,
\label{A2corrform2}
\end{eqnarray}
which exactly corresponds to the divergence (\ref{a2a4a6m0}). Thus, indeed, the divergences arising in the massless theory are entirely determined by the zero Matsubara modes.

The regularization proposed in the Section \ref{sec:real}, in fact, has serious physical grounds. We consider integrals independent of the dimensional parameter, which, by definition, satisfy the relation \cite{Pak:2010pt}
\begin{eqnarray}
I(\alpha p)=\alpha^{l}I(p)\,,
\label{freesc}
\end{eqnarray}
in particular, $ \int d (\alpha|{\bf p}|)/ (\alpha|{\bf p}|)^{2n}=\alpha^{1-2n} \int d|{\bf p}|/ |{\bf p}|^{2n} $. It turns out that in dimensional regularization, such integrals are always zero
\begin{eqnarray}
I(p)=0\,,
\label{freesc1}
\end{eqnarray}
which is connected with the axioms and definitions of dimensional regularization \cite{Collins:1984xc, Pak:2010pt} and is sometimes called Veltman's formula in the literature \cite{Kleinert:2001ax}. Property (\ref{freesc1}) is often used in quantum field theory, including quantum field theory at finite temperatures (see, in particular, Chapter 2 of \cite{Laine:2016hma}).

It would seem that in this way, we immediately get confirmation of the recipe for dealing with these divergences by subtracting them. However, as noted, for example, in \cite{Laine:2016hma}, the situation becomes nontrivial, if higher orders of perturbation theory lead to an infinite series of similar divergences. And then, despite the fact that each of the diverging integrals separately is axiomatically zero, their sum can give a finite contribution.

To test the possibility of such a situation, it is necessary to consider higher orders of perturbation theory in acceleration, which is a difficult task from a technical point of view. However, a partial answer to the question about the divergences in higher orders can be obtained if we consider only the contribution of zero modes.

In particular, the sixth-order term in perturbation theory is described by the formula
\begin{eqnarray}
&&\rho\sim A_3 |a|^{6}=\int \frac{d\tau_1... d\tau_6 d^3 x_1... d^3 x_6 d^3 p_1 ...d^3 p_7}{6!(2\pi)^{18}}x_1^{(3)}...x_6^{(3)} \nonumber  \\
&&\times e^{i{\bf p}_1{\bf x}_1-i{\bf p}_2({\bf x}_1-{\bf x}_2)-i{\bf p}_3({\bf x}_2-{\bf x}_3)-i{\bf p}_4({\bf x}_3-{\bf x}_4)-i{\bf p}_5({\bf x}_4-{\bf x}_5)-i{\bf p}_6({\bf x}_5-{\bf x}_6)-i{\bf p}_7{\bf x}_6}\nonumber  \\
&&
\times \sum_{\omega_{1n}...\omega_{7n}}\frac{1}{|\beta|^7} e^{-i\omega_{1n}\tau_{1}+i\omega_{2n}(\tau_{1}-\tau_{2})+i\omega_{3n}(\tau_{2}-\tau_{3})+i\omega_{4n}(\tau_{3}-\tau_{4})+i\omega_{5n}(\tau_{4}-\tau_{5})+i\omega_{6n}(\tau_{5}-\tau_{6})+i\omega_{7n}\tau_{6}}\nonumber \\
&&\times \Delta(P_1)\Delta(P_2)\Delta(P_3)\Delta(P_4)\Delta(P_5)\Delta(P_6)\Delta(P_7)\nonumber \\
&&\times\mathcal{D}^{00}(iP_1 ,-iP_2 )
\mathcal{D}^{00}(iP_2 ,-iP_3 )
\mathcal{D}^{00}(iP_3 ,-iP_4 )
\mathcal{D}^{00}(iP_4 ,-iP_5 )
\mathcal{D}^{00}(iP_5 ,-iP_6 )\nonumber \\
&&\times\mathcal{D}^{00}(iP_6 ,-iP_7 )
\mathcal{D}^{00}(iP_7 ,-iP_1 )+.... \,,
\label{a6}
\end{eqnarray}
where the ellipsis corresponds to the contribution of the other terms coming from the Wick theorem. Using the equation
\begin{eqnarray}
&&\int d^3 p_1 ...d^3 p_7 d^3 x_1... d^3 x_6  \,F({\bf p}_1,...,{\bf p}_7) x_1^{(3)}...x_6^{(3)} \nonumber \\
&&e^{i{\bf p}_1{\bf x}_1-i{\bf p}_2({\bf x}_1-{\bf x}_2)-i{\bf p}_3({\bf x}_2-{\bf x}_3)-i{\bf p}_4({\bf x}_3-{\bf x}_4)-i{\bf p}_5({\bf x}_4-{\bf x}_5)-i{\bf p}_6({\bf x}_5-{\bf x}_6)-i{\bf p}_7{\bf x}_6}  \nonumber \\
&&=(2\pi)^{18}\int d^3p \frac{\partial^2}{\partial p_1^{(3)}\partial p_7^{(3)}}
\Big(\frac{\partial}{\partial p_2^{(3)}}
+\frac{\partial}{\partial p_1^{(3)}}\Big)...
\nonumber \\
&&\times\Big(\frac{\partial}{\partial p_5^{(3)}}
+...+\frac{\partial}{\partial p_1^{(3)}}\Big)F({\bf p}_1,...,{\bf p}_7) \Big|_
{\scriptstyle {\bf p}_i={\bf p}}\,,
\label{dif re1l}
\end{eqnarray}
and considering the contribution of the zero modes $ \omega_1=...=\omega_7=0 $ we get that the contribution of the correlator under consideration is equal to zero. Assuming that, as in the case of fourth-order coefficients, the contributions of zero modes from different correlators arising from Wick theorem are equal, we obtain
\begin{eqnarray}
A_{3}^{\omega=0} |a|^{6}=0 \,.
\label{a61}
\end{eqnarray}
In the same way we also obtain
\begin{eqnarray}
A_{4}^{\omega=0} |a|^{8}=0 \,.
\label{a8}
\end{eqnarray}
Thus, at least, the leading divergences in the first two higher orders of the perturbation theory apparently do not arise. If this situation is general (no leading divergences in higher orders and no subleading divergences), we can remove the divergences from the expressions (\ref{a2a4a6m0}) based on (\ref{freesc1}).

We now consider the divergences for the finite mass (\ref{a1a3a5m2}) and (\ref{a2a4a6m2}). For a finite mass, all divergences arising, both $ \int d|{\bf p}|/|{\bf p}|^{2} $ and $ \int d|{\bf p}|/|{\bf p}|^{4} $, are fictitious in the sense that the integral formulas themselves, in particular (\ref{coef old 1}), as functions of mass, do not contain divergences, which can be verified directly numerically.

On the other hand, if we sum up all the divergences in all orders in mass, then, as we will now show, we eliminate the fictitious divergences and get the correct expression for the perturbative series. A similar procedure of summation of infrared divergences was realized, in particular, in \cite{Laine:2016hma}.

We first consider the divergences in the order $ a^2 $. By directly decomposing the integrand in (\ref{coef old 1}) and isolating the divergences as negative-power terms of the Laurent series, we can show that a series of divergences arises
\begin{eqnarray}
A_{1}&=&\frac{1}{12}+\frac{1}{96\pi^2}\frac{m^2}{T^2}-
\frac{1}{8\pi^2 T}\int_0^{\infty}d|{\bf p}|\frac{m^2}{|{\bf p}|^2}
+
\frac{1}{8\pi^2 T}\int_0^{\infty}d|{\bf p}|\frac{m^4}{|{\bf p}|^4}\nonumber \\
&&-
\frac{1}{8\pi^2 T}\int_0^{\infty}d|{\bf p}|\frac{m^6}{|{\bf p}|^6}+... \,,
\label{a1infr}
\end{eqnarray}
where the ellipsis includes both terms of higher orders and integrals from the nondivergent parts of the orders $ m^4, m^6 $. In this case, there are no subleading infrared divergences. We can also assume that the infrared divergences form a geometric progression. To sum up these divergences mathematically rigorously, we consider the integrand $ A_1(m)=\int_0^{\infty}d|{\bf p}|a_1(m^2,|{\bf p}|^2) $ in (\ref{coef old 1}). It is easy to prove that by expanding it in a series of $ m^2 $, the leading divergence in the order $ m^{2n} $ has the form $ m^{2n}/|{\bf p}|^{2n} $. Then the expansion of $ a_1(m^2,|{\bf p}|^2) $ in a series has the form
\begin{eqnarray}
a_1(m,|{\bf p}|)=a_1(0,|{\bf p}|)+\sum_{n=1}^{\infty} g_n \frac{m^{2n}}{|{\bf p}|^{2n}}+\sum_{n=1}^{\infty} g^{(1)}_n\frac{m^{2n}}{|{\bf p}|^{2n-1}}+...  \,.
\label{a1infr1}
\end{eqnarray}
We will replace $ |{\bf p}|\to\lambda|{\bf p}|,\,m\to\lambda m $, after which we find the limit $ \lambda\to 0 $. This will allow us to sum up the leading infrared divergences
\begin{eqnarray}
\lim_{\lambda\to 0}a_1(\lambda^2 m^2, \lambda^2|{\bf p}|^2)-a_1(0,\lambda^2|{\bf p}|^2)=\sum_{n=1}^{\infty} g_n \frac{m^{2n}}{|{\bf p}|^{2n}}\,.
\label{a1infr2}
\end{eqnarray}
The left-hand side in (\ref{a1infr2}) can be easily found, we get
\begin{eqnarray}
-\frac{1}{8\pi^2 T}\frac{m^2}{m^2+|{\bf p}|^2}=\sum_{n=1}^{\infty} g_n \frac{m^{2n}}{|{\bf p}|^{2n}}\,,
\label{a1infr3}
\end{eqnarray}
and, thus, the assumption made in (\ref{a1infr}), that the leading infrared divergences form a geometric progression, is confirmed. In particular, expanding the left-hand side of (\ref{a1infr3}) in a series, we would obtain the terms (\ref{a1infr}).

Integrating (\ref{a1infr3}), we obtain a contribution to the energy density
\begin{eqnarray}
\int_0^{\infty}d|{\bf p}|\sum_{n=1}^{\infty} g_n \frac{m^{2n}}{|{\bf p}|^{2n}}=-\int_0^{\infty}d|{\bf p}|\frac{1}{8\pi^2 T}\frac{m^2}{m^2+|{\bf p}|^2}=-\frac{m}{16\pi T}  \,.
\label{a1infr4}
\end{eqnarray}

The contribution (\ref{a1infr4}) can be called nonperturbative in mass, since it has the form $ \sqrt{m^2} $ and does not directly follow from the perturbative expansion in $ m^2 $. Thus, in the order $ m^2 $, taking into account the contribution of infrared divergences, we obtain
\begin{eqnarray}
A_{1}=\frac{1}{12}+\frac{1}{96\pi^2}\frac{m^2}{T^2}-\frac{m}{16\pi T}+... \,,
\label{a1infr5}
\end{eqnarray}
where the ellipsis denotes the potential contribution of subleading divergences and finite terms in orders $ m^{2n}, n>1 $.

Note that the correctness of the result obtained (\ref{a1infr5}) can be checked numerically by finding the numerical value of the derivatives of the integral $ \frac{\partial }{\partial m} A_1(m)$ and $ \frac{\partial^2 }{\partial m^2} A_1(m)$, which, in particular, are finite.

We can also rigorously prove that there are no subleading divergences. In particular, this follows from the equality
\begin{eqnarray}
\sum_{n=1}^{\infty} g^{(k)}_n \frac{m^{2n}}{|{\bf p}|^{2n-k}}=\lim_{\lambda\to 0}\frac{\partial^{k}}{k!\partial \lambda^{k}}\Big(a_1(\lambda^2 m^2, \lambda^2|{\bf p}|^2)-a_1(0,\lambda^2|{\bf p}|^2)\Big)\,.
\label{a1infr6}
\end{eqnarray}
The absence of subleading divergences  corresponds to equality of (\ref{a1infr6}) to a finite polynomial with positive powers. This has been  verified for a number of values $ k $ ($ k=1,2,3,4 $).

A similar situation will be with the divergences in the order $ a^4 $. In particular, the divergence $ \int d|{\bf p}|/|{\bf p}|^{2} $, which, as previously described, can be excluded in the massless case, since it is equal to zero, should now be summed with all terms of higher orders $ m^{2n} $. There is no contradiction with what we talked about previously in the massless case, since each of the divergences is equal to zero separately, but their sum can be finite.

First, we need an expression for the coefficient in energy density for an arbitrary mass, without expansion in mass. This can be done by holding the mass in the energy-momentum tensor and propagators, as a result of which we obtain an expression similar to (\ref{A2Int})
\begin{eqnarray}
A_2(m) &=& \int_0^{\infty} \frac{d |{\bf p}|}{72\pi^2}\Bigg(\Big[\frac{3m^2|{\bf p}|^2}{4E_p}+\frac{|{\bf p}|^4}{E_p}\Big]n_B^{(4)}(E_p)+
\Big[\frac{3m^2|{\bf p}|^2}{10}+|{\bf p}|^4\Big]n_B^{(5)}(E_p)\nonumber \\
&&+\Big[\frac{m^4|{\bf p}|^2}{20E_p}+\frac{13m^2|{\bf p}|^4}{40E_p}+\frac{7|{\bf p}|^6}{20E_p}\Big] n_B^{(6)}(E_p)+
\Big[\frac{m^2|{\bf p}|^4}{70}+\frac{|{\bf p}|^6}{25}\Big] n_B^{(7)}(E_p) \nonumber \\
&&+\frac{9 E_p |{\bf p}|^6}{5600} n_B^{(8)}(E_p)\Bigg)
\,.
\label{A2Intm}
\end{eqnarray}
Expanding (\ref{A2Intm}) in a series in mass, we obtain a series of divergences
\begin{eqnarray}
A_{2}&=&-\frac{11}{480\pi^2}+
\frac{4 T}{15\pi^2}\int_0^{\infty}d|{\bf p}|\frac{1}{|{\bf p}|^2}
-
\frac{2 T}{5\pi^2}\int_0^{\infty}d|{\bf p}|\frac{m^2}{|{\bf p}|^4}+
\frac{T}{4\pi^2}\int_0^{\infty}d|{\bf p}|\frac{m^4}{|{\bf p}|^6} \nonumber \\
&&+\frac{T}{3\pi^2}\int_0^{\infty}d|{\bf p}|\frac{m^6}{|{\bf p}|^8}+... \,.
\label{a2infr}
\end{eqnarray}
As in the previous case, we see that there are no subleading divergences. It can be shown that the leading divergences are of the order $ m^{2n-2}/|{\bf p}|^{2n} $. Accordingly, expanding the integrand in (\ref{a2infr}), we obtain
\begin{eqnarray}
a_2(m^2,|{\bf p}|^2)=\sum_{n=1}^{\infty} g_n \frac{m^{2n-2}}{|{\bf p}|^{2n}}+\sum_{n=1}^{\infty} g^{(1)}_n\frac{m^{2n-2}}{|{\bf p}|^{2n-1}}+...  \,.
\label{a2infr1}
\end{eqnarray}
Replacing $ |{\bf p}|\to\lambda|{\bf p}|,\,m\to\lambda m $ and taking the limit $ \lambda\to 0 $, we get
\begin{eqnarray}
\sum_{n=1}^{\infty} g_n \frac{m^{2n-2}}{|{\bf p}|^{2n}}=\lim_{\lambda\to 0}\lambda^2 a_2(\lambda^2 m^2, \lambda^2|{\bf p}|^2)=T\frac{|{\bf p}|^2(15 m^4+40 m^2 |{\bf p}|^2+16 |{\bf p}|^4)}{60 \pi^2(m^2+|{\bf p}|^2)^4}\,.
\label{a2infr2}
\end{eqnarray}
Thus, we have found the sum of the leading divergences.

After integration we obtain
\begin{eqnarray}
\int_0^{\infty}d|{\bf p}|\sum_{n=1}^{\infty} g_n \frac{m^{2n-2}}{|{\bf p}|^{2n}}=\frac{9 T}{128 \pi m}  \,,
\label{a2infr3}
\end{eqnarray}
and the contribution to the coefficient $ A_2 $
\begin{eqnarray}
A_{2}=-\frac{11}{480 \pi^2}+\frac{9 T}{128 \pi m} +... \,,
\label{a2infr4}
\end{eqnarray}
which is also confirmed numerically.

Thus, the infrared divergences found in the Section \ref{sec:real} can indeed be excluded both in the massless case and for the massive theory. In the case of a massive theory, these divergences can be summed up and give a finite nonperturbative and odd-in-mass contribution, and in this sense they can be excluded from terms of the zeroth and $ m^2 $ order. The study of these terms from the point of view of the Unruh effect requires additional consideration of nonperturbative in acceleration terms, which is beyond the scope of this paper.

We also note that there is another way to cancel the infrared divergences, if we require the execution of $ T^{\mu\nu}(T=T_U)=0 $ from the very beginning. This approach is similar to \cite{Frolov:1987dz, Dowker:1987pk, Bezerra:2006nu} and also leads to the exclusion of terms with divergences. However, nonperturbative terms $ m |a|^2, |a|^4/m $ may be lost.

In conclusion, we discuss a possible connection with Bose condensation. Indeed, according to, for example, \cite{Laine:2016hma}, the appearance of infrared divergences is a possible indication of Bose condensation. The question of the possibility of Bose condensation due to acceleration is a non-trivial problem requiring a separate consideration. The main argument in favor of the absence of such a phenomenon is the lack of such condensate in the dual approach with a conical singularity. In this approach, condensation does not occur, at least in the temperature range $ T>T_U $.

In the statement of the problem, we consider a gas of free particles. At zero chemical potential, Bose condensate does not occur in such a gas. When the system acquires acceleration, we would not expect the appearance of condensate, since acceleration leads to heating of the medium, and temperature, as a rule, plays the role of a factor that destroys the condensate.

The conclusion is that the appearance of infrared divergences is typical consequence of the Bose distribution, but is not necessarily associated with Bose condensation (in particular, in the simplest example, expanding the energy density of the Bose gas in a series in mass, we would also get a similar series of infrared divergences).

\section{Discussion}
\label{sec:disc}

In the Sections \ref{sec:real}, \ref{sec:complex}, \ref{sec:fermi} we calculated quantum corrections related to acceleration to the energy-momentum tensor. We have shown that both in the case of massless fields and in the case of massive fields in the first order in mass, the calculated corrections satisfy the condition following from the Unruh effect. It is clear, since the calculated mean value of the energy-momentum tensor is normalized to the Minkowski vacuum, which corresponds to taking all the products of the operators using normal ordering  \cite{Becattini:2015nva}. Then, according to the Unruh effect, the Minkowski vacuum is perceived by the accelerated observer as a heat bath with Unruh temperature. Therefore, at the proper temperature equal to the Unruh temperature, the mean value of the energy-momentum tensor should be zero. This is what we have shown for a wide class of theories. Thus, we can talk about confirming the universality of the Unruh effect in the framework of the Zubarev approach, which is a general quantum-field phenomenon that does not depend on the type of fields under consideration.

We have also shown the emergence of conical geometry \cite{Dowker:1994fi, Dowker:1987pk, Iellici:1998ce, Bezerra:2006nu, Frolov:1987dz, Iellici:1997ud} in the statistical Zubarev approach \cite{Zubarev, Buzzegoli:2017cqy,Weert}. The expressions obtained in the framework of statistical approach exactly coincide with the mean values calculated in the space-time with a conical singularity both in the case of massless fields and in the quadratic order in mass.

In particular, this allows one to obtain expressions for an accelerated medium from known results for cosmic strings and vice versa.

An important consequence about the non-renormalization of the obtained expressions follows from this duality.  When considering the formulas (\ref{tensorrealm0}), (\ref{tensorrealm2}), (\ref{tensorcomplexm0}), (\ref{tensorcomplexm2}), (\ref{tensorfermim0}), (\ref{tensorfermim2}) the question arises as to whether the calculated acceleration orders are maximal. Now we have received a positive answer to this question. Indeed, in formulas (\ref{enconm0a}), (\ref{enconm2scal}) and (\ref{bez6}), obtained using the geometrical approach, the calculated orders are maximal as the acceleration effects are taken into account in a nonperturbative way.

Thus, we can make a prediction that the calculation of subsequent acceleration corrections, in which negative degrees of temperature could formally appear, within the framework of the statistical approach with the operator (\ref{operk}) will give zero. We have also verified this statement directly in a particular case of corrections of the order $ m^2 |a|^4/T^2 $ in the coefficients $ A_2^{m2}, A_4^{m2}$ and $A_6^{m2} $ in (\ref{a2a4a6m2}) and (\ref{coeffermim2}). These corrections turned out to be zero (in the case of scalar fields, after subtracting the divergences).

We also note that absence of higher orders or polynomiality is associated with the properties of Sommerfeld integrals when integrated in the complex plane \cite{Prokhorov:2019hif} as was also discussed in Section \ref{sec:intro2}. Moreover, according to  \cite{Prokhorov:2019hif}, it is necessary to make a substantial remark that formulas of the form  (\ref{tensorrealm0}), (\ref{tensorrealm2}), (\ref{tensorcomplexm0}), (\ref{tensorcomplexm2}), (\ref{tensorfermim0}), (\ref{tensorfermim2}) are exact nonperturbative expressions only in the region $ T>T_U $. When considering the domain $ T<T_U $, perturbative formulas of the form (\ref{tensorfermim0}) may stop to be applicable and additional nonperturbative contributions can appear. Moreover, we can talk about the existence of instability at the boundary of two regions at $ T=T_U $ at least for Dirac field.

\section{Conclusions}
\label{sec:concl}

We have calculated quantum corrections related to the acceleration to free-field energy-momentum tensors using the statistical Zubarev density operator. A wide class of theories is considered: massless real and complex scalar fields, Dirac field, and also massive fields in the quadratic order in mass. All calculated corrections satisfy the Unruh effect: the energy-momentum tensor, taking into account the obtained corrections, turns out to be zero at the proper temperature equal to the Unruh temperature. Thus, the universality of the Unruh effect in the statistical approach is demonstrated. 

In the case of Dirac field, the studied corrections lead to finite momentum integrals, both in the massless case and for the corrections of the order of $ m^2 $. However, for scalar fields, infrared divergences appear in the corrections of the order $ \mathcal{O}(|a|^4) $ and $ \mathcal{O}(m^2 |a|^2) $. In the case of massless theory, these divergences correspond to the Matsubara zero modes and can be excluded based on Veltman's formula. In the case of massive fields, a summation of an infinite series of infrared divergences can be made, as a result of which it was shown that they contribute to the odd terms in mass, a detailed study of which is beyond the scope of this paper. To summarize, effectively these divergences can be regularized by subtracting the corresponding negative power terms of the Laurent series of the form $ 1/|{\bf p}|^2 $ and $ 1/|{\bf p}|^4 $. As a result, we obtain the expressions for the energy-momentum tensor satisfying the Unruh effect.

It is shown that in the used statistical approach of Zubarev, conical geometry emerges.
In particular, the calculated quantum corrections exactly correspond to the corrections calculated in the framework of field theory in a space with a conical singularity \cite{Dowker:1994fi, Dowker:1987pk, Iellici:1998ce, Bezerra:2006nu, Frolov:1987dz, Iellici:1997ud} in all cases considered. We can also talk about the duality of statistical and geometrical approaches. This duality was first noted in \cite{Prokhorov:2019hif}, and in this paper we show that it is a general phenomenon.

We began to study the consequences of the discovered duality. In particular, since all the expressions calculated in a space with a conical singularity were exact and nonperturbative, it should be expected that in the statistical approach all the higher order terms will be also equal to zero. We have verified this directly in particular cases by calculating quantum corrections of the order of $ m^2 |a|^4/T^2 $ for all considered field theories. All corrections of this type turned out to be equal to zero.

The statistical approach inherently describes an effective interaction, introduced on a macroscopic scale or in the infrared. At the same time, the dual quantum field approach in a space with a conical singularity, as expected, on the contrary should be valid on a microscopic scale in the ultraviolet region. The existence of the duality of these two approaches makes it possible to smoothly join two regions of scales.

In addition, duality allows us to discuss the issue of Bose condensation in an accelerated medium. From the point of view of the statistical approach, Bose condensation associated with acceleration should be investigated as a separate non-trivial case. On the other hand, field theory with a conical singularity indicates the absence of such a phenomenon, which is confirmed by general arguments about the relationship between acceleration and temperature resulting from the Unruh effect.

{\bf Acknowledgements}

We are grateful to F. Becattini, M. Bordag and D. Fursaev for useful discussions and comments.
The reported study was funded by RFBR according to the research projects 18-02-40056 (Prokhorov),
17-02-01108 (Teryaev), 18-02-40084 (Zakharov).



\begin{thebibliography}{99}

\bibitem{Kharzeev:2012ph}
  D.~E.~Kharzeev, K.~Landsteiner, A.~Schmitt and H.~U.~Yee,
  \emph{'Strongly interacting matter in magnetic fields': an overview},\emph{
  Lect.\ Notes Phys.\ }  {\bf 871} (2013) doi:10.1007/978-3-642-37305-3\_1
  [arXiv:1211.6245 [hep-ph]].

\bibitem{Son:2009tf}
  D.~T.~Son and P.~Surowka,
  \emph{Hydrodynamics with Triangle Anomalies},\emph{
  Phys.\ Rev.\ Lett.\ }  {\bf 103}, 191601 (2009)
  doi:10.1103/PhysRevLett.103.191601
  [arXiv:0906.5044 [hep-th]].

\bibitem{Vilenkin:1980zv}
  A.~Vilenkin,
  \emph{Quantum Field Theory At Finite Temperature In A Rotating System},\emph{
  Phys.\ Rev.\ D } {\bf 21}, 2260 (1980).
  doi:10.1103/PhysRevD.21.2260.

\bibitem{Sadofyev:2010is}
  A.~V.~Sadofyev, V.~I.~Shevchenko and V.~I.~Zakharov,
  \emph{Notes on chiral hydrodynamics within effective theory approach},\emph{
  Phys.\ Rev.\ D } {\bf 83}, 105025 (2011)
  doi:10.1103/PhysRevD.83.105025
  [arXiv:1012.1958 [hep-th]].

\bibitem{Stone:2018zel}
  M.~Stone and J.~Kim,
  \emph{Mixed Anomalies: Chiral Vortical Effect and the Sommerfeld Expansion},\emph{
  Phys.\ Rev.\ D } {\bf 98}, no. 2, 025012 (2018)
  doi:10.1103/PhysRevD.98.025012
  [arXiv:1804.08668 [cond-mat.mes-hall]].

\bibitem{Unruh:1976db}
  W.~G.~Unruh,
  \emph{Notes on black hole evaporation},\emph{
  Phys.\ Rev.\ D } {\bf 14}, 870 (1976).
  doi:10.1103/PhysRevD.14.870

\bibitem{Jiang:2016wvv}
  Y.~Jiang and J.~Liao,
  \emph{Pairing Phase Transitions of Matter under Rotation},\emph{
  Phys.\ Rev.\ Lett.\ }  {\bf 117}, no. 19, 192302 (2016)
  doi:10.1103/PhysRevLett.117.192302
  [arXiv:1606.03808 [hep-ph]].

\bibitem{Castorina:2012yg}
  P.~Castorina and M.~Finocchiaro,
  \emph{Symmetry Restoration By Acceleration},\emph{
  J.\ Mod.\ Phys.\ }  {\bf 3}, 1703 (2012)
  doi:10.4236/jmp.2012.311209
  [arXiv:1207.3677 [hep-th]].

\bibitem{Ohsaku:2004rv}
  T.~Ohsaku,
  \emph{Dynamical chiral symmetry breaking and its restoration for an accelerated observer},\emph{
  Phys.\ Lett.\ B } {\bf 599}, 102 (2004)
  doi:10.1016/j.physletb.2004.08.019
  [hep-th/0407067].

\bibitem{Takeuchi:2015nga}
  S.~Takeuchi,
  \emph{Bose-Einstein condensation in the Rindler space},\emph{
  Phys.\ Lett.\ B } {\bf 750}, 209 (2015)
  doi:10.1016/j.physletb.2015.09.013
  [arXiv:1501.07471 [hep-th]].

\bibitem{Rogachevsky:2010ys}
  O.~Rogachevsky, A.~Sorin and O.~Teryaev,
  \emph{Chiral vortaic effect and neutron asymmetries in heavy-ion collisions},\emph{
  Phys.\ Rev.\ C } {\bf 82} (2010) 054910
  doi:10.1103/PhysRevC.82.054910
  [arXiv:1006.1331 [hep-ph]].

\bibitem{Florkowski:2018fap}
  W.~Florkowski, R.~Ryblewski and A.~Kumar,
  \emph{Relativistic hydrodynamics for spin-polarized fluids},\emph{
  Prog.\ Part.\ Nucl.\ Phys.\ }  {\bf 108}, 103709 (2019)
  doi:10.1016/j.ppnp.2019.07.001
  [arXiv:1811.04409 [nucl-th]].

\bibitem{Becattini:2019ntv}
  F.~Becattini, G.~Cao and E.~Speranza,
  \emph{Polarization transfer in hyperon decays and its effect in relativistic nuclear collisions},\emph{
  Eur.\ Phys.\ J.\ C } {\bf 79}, no. 9, 741 (2019)
  doi:10.1140/epjc/s10052-019-7213-6
  [arXiv:1905.03123 [nucl-th]].

\bibitem{Baznat:2017jfj}
  M.~Baznat, K.~Gudima, A.~Sorin and O.~Teryaev,
  \emph{Hyperon polarization in heavy-ion collisions and holographic gravitational anomaly},\emph{
  Phys.\ Rev.\ C } {\bf 97}, no. 4, 041902 (2018)
  doi:10.1103/PhysRevC.97.041902
  [arXiv:1701.00923 [nucl-th]].

\bibitem{Becattini:2008tx}
  F.~Becattini, P.~Castorina, J.~Manninen and H.~Satz,
  \emph{The Thermal Production of Strange and Non-Strange Hadrons in e+ e- Collisions},\emph{
  Eur.\ Phys.\ J.\ C } {\bf 56}, 493 (2008)
  doi:10.1140/epjc/s10052-008-0671-x
  [arXiv:0805.0964 [hep-ph]].

\bibitem{Prokhorov:2019hif}
G.~Y.~Prokhorov, O.~V.~Teryaev and V.~I.~Zakharov,
\emph{Thermodynamics of accelerated fermion gases and their instability at the Unruh temperature} , \emph{
Phys.\ Rev.\ D} {\bf 100}, no. 12, 125009 (2019)
doi:10.1103/PhysRevD.100.125009
[arXiv:1906.03529 [hep-th]].

\bibitem{Buzzegoli:2017cqy}
  M.~Buzzegoli, E.~Grossi and F.~Becattini,
  \emph{General equilibrium second-order hydrodynamic coefficients for free quantum fields},\emph{
  JHEP } {\bf 1710} (2017) 091
  doi:10.1007/JHEP10(2017)091
  [arXiv:1704.02808 [hep-th]].

\bibitem{Buzzegoli:2018wpy}
  M.~Buzzegoli and F.~Becattini,
  \emph{General thermodynamic equilibrium with axial chemical potential for the free Dirac field},\emph{
  JHEP } {\bf 1812}, 002 (2018)
  doi:10.1007/JHEP12(2018)002
  [arXiv:1807.02071 [hep-th]].

\bibitem{Gao:2012ix}
  J.~H.~Gao, Z.~T.~Liang, S.~Pu, Q.~Wang and X.~N.~Wang,
  \emph{Chiral Anomaly and Local Polarization Effect from Quantum Kinetic Approach},\emph{
  Phys.\ Rev.\ Lett.\ }  {\bf 109}, 232301 (2012)
  doi:10.1103/PhysRevLett.109.232301
  [arXiv:1203.0725 [hep-ph]].

\bibitem{Prokhorov:2018qhq}
  G.~Prokhorov, O.~Teryaev and V.~Zakharov,
  \emph{Axial current in rotating and accelerating medium},\emph{
  Phys.\ Rev.\ D } {\bf 98}, no. 7, 071901 (2018)
  doi:10.1103/PhysRevD.98.071901
  [arXiv:1805.12029 [hep-th]].

\bibitem{Prokhorov:2018bql}
  G.~Y.~Prokhorov, O.~V.~Teryaev and V.~I.~Zakharov,
  \emph{Effects of rotation and acceleration in the axial current: density operator vs Wigner function},\emph{
  JHEP } {\bf 1902}, 146 (2019)
  doi:10.1007/JHEP02(2019)146
  [arXiv:1807.03584 [hep-th]].

\bibitem{Fukushima:2008xe}
  K.~Fukushima, D.~E.~Kharzeev and H.~J.~Warringa,
  \emph{The Chiral Magnetic Effect},\emph{
  Phys.\ Rev.\ D } {\bf 78}, 074033 (2008)
  doi:10.1103/PhysRevD.78.074033
  [arXiv:0808.3382 [hep-ph]].

\bibitem{Zubarev}
D. N. Zubarev, A. V. Prozorkevich, S. A. Smolyanskii, \emph{Derivation of nonlinear generalized equations of quantum relativistic hydrodynamics} , \emph{TMF}, 40:3 (1979), 394-407; \emph{Theoret. and Math. Phys.}, 40:3 (1979), 821-831.

\bibitem{Weert}
G. Van Weert, \emph{Maximum entropy principle and relativistic hydrodynamics}, \emph{Ch. Annals Phys.}, Volume 140, Issue 1, (1982), 133-162

\bibitem{Becattini:2019dxo}
  F.~Becattini, M.~Buzzegoli and E.~Grossi,
  \emph{Reworking the Zubarev's approach to non-equilibrium quantum statistical mechanics},
  \emph{Particles} {\bf 2}, no. 2, 197 (2019)
  doi:10.3390/particles2020014
  [arXiv:1902.01089 [cond-mat.stat-mech]].

\bibitem{Becattini:2015nva}
  F.~Becattini and E.~Grossi,
  \emph{Quantum corrections to the stress-energy tensor in thermodynamic equilibrium with acceleration},\emph{
  Phys.\ Rev.\ D } {\bf 92}, 045037 (2015)
  doi:10.1103/PhysRevD.92.045037
  [arXiv:1505.07760 [gr-qc]].

\bibitem{Prokhorov:2019cik}
  G.~Y.~Prokhorov, O.~V.~Teryaev and V.~I.~Zakharov,
  \emph{Unruh effect for fermions from the Zubarev density operator},\emph{
  Phys.\ Rev.\ D } {\bf 99}, no. 7, 071901 (2019)
  doi:10.1103/PhysRevD.99.071901
  [arXiv:1903.09697 [hep-th]].

\bibitem{Becattini:2017ljh}
  F.~Becattini,
  \emph{Thermodynamic equilibrium with acceleration and the Unruh effect},\emph{
  Phys.\ Rev.\ D } {\bf 97}, no. 8, 085013 (2018)
  doi:10.1103/PhysRevD.97.085013
  [arXiv:1712.08031 [gr-qc]].

\bibitem{Frolov:1987dz}
  V.~P.~Frolov and E.~M.~Serebryanyi,
  \emph{Vacuum Polarization in the Gravitational Field of a Cosmic String},\emph{
  Phys.\ Rev.\ D } {\bf 35}, 3779 (1987).
  doi:10.1103/PhysRevD.35.3779

\bibitem{Dowker:1994fi}
  J.~S.~Dowker,
  \emph{Remarks on geometric entropy},\emph{
  Class.\ Quant.\ Grav.\ }  {\bf 11}, L55 (1994)
  doi:10.1088/0264-9381/11/4/001
  [hep-th/9401159].


\bibitem{Florkowski:2018myy}
  W.~Florkowski, E.~Speranza and F.~Becattini,
  \emph{Perfect-fluid hydrodynamics with constant acceleration along the stream lines and spin polarization},\emph{
  Acta Phys.\ Polon.\ B } {\bf 49}, 1409 (2018)
  doi:10.5506/APhysPolB.49.1409
  [arXiv:1803.11098 [nucl-th]].

\bibitem{Dowker:1987pk}
  J.~S.~Dowker,
  \emph{Vacuum Averages for Arbitrary Spin Around a Cosmic String},\emph{
  Phys.\ Rev.\ D } {\bf 36}, 3742 (1987).
  doi:10.1103/PhysRevD.36.3742

\bibitem{Iellici:1998ce}
  D.~Iellici,
  \emph{Aspects and applications of quantum field theory on spaces with conical singularities},
  gr-qc/9805058.

\bibitem{Iellici:1997ud}
  D.~Iellici,
  \emph{Massive scalar field near a cosmic string},
  \emph{Class.\ Quant.\ Grav.\ } {\bf 14}, 3287 (1997)
  doi:10.1088/0264-9381/14/12/013
  [gr-qc/9704077].

\bibitem{Bezerra:2006nu}
  V.~B.~Bezerra and N.~R.~Khusnutdinov,
  \emph{Vacuum expectation value of the spinor massive field in the cosmic string space-time},
  \emph{Class.\ Quant.\ Grav.\ } {\bf 23}, 3449 (2006)
  doi:10.1088/0264-9381/23/10/015
  [hep-th/0602048].

 \bibitem{landau}
 Landau L. D., Lifshitz E. M., \emph{Statistical Physics}, (2013) Elsevier.


 \bibitem{Prokhorov:2019sss}
 G.~Y.~Prokhorov, O.~V.~Teryaev and V.~I.~Zakharov,
 \emph{Calculation of acceleration effects using the Zubarev density operator} ,\emph{
 	Particles} {\bf 3}, no. 1, 1 (2020)
 doi:10.3390/particles3010001
 [arXiv:1911.04563 [hep-th]].

\bibitem{Collins:1984xc}
J.~C.~Collins,
\emph{Renormalization : An Introduction to Renormalization, The Renormalization Group, and the Operator Product Expansion},
doi:10.1017/CBO9780511622656

\bibitem{Kleinert:2001ax}
H.~Kleinert and V.~Schulte-Frohlinde,
\emph{Critical properties of phi**4-theories},
River Edge, USA: World Scientific (2001) 489 p

\bibitem{Pak:2010pt}
A.~Pak and A.~Smirnov,
\emph{Geometric approach to asymptotic expansion of Feynman integrals},
\emph{Eur.\ Phys.\ J.\ C} {\bf 71}, 1626 (2011)
doi:10.1140/epjc/s10052-011-1626-1
[arXiv:1011.4863 [hep-ph]].

\bibitem{Laine:2016hma}
  M.~Laine and A.~Vuorinen,
  \emph{Basics of Thermal Field Theory},
  \emph{Lect.\ Notes Phys.\ }  {\bf 925}, pp.1 (2016)
  doi:10.1007/978-3-319-31933-9
  [arXiv:1701.01554 [hep-ph]].

\end{thebibliography}
\end{document}